\documentclass[journal,twoside]{IEEEtran}
\newcommand{\qeda}{\hfill\ensuremath{\blacksquare}}
\newcommand{\f}{it follows that }

\newcommand{\m}[1]{\mbox{$#1$}}

\ifCLASSINFOpdf
\else
\fi

\usepackage{cite}
\usepackage{amsfonts}
\usepackage{mathrsfs}
\usepackage{csquotes}
\usepackage{bbm}
\usepackage{pifont}
\newcommand{\argmin}{\operatornamewithlimits{argmin}}
\usepackage{amsfonts}
\usepackage{amsmath, amsthm}
\usepackage{tabularx}
\usepackage{listings}
\usepackage{graphicx}
\usepackage{float}
\usepackage{amssymb}
\usepackage{array}
\usepackage{fancyhdr}
\usepackage{cases}
 \usepackage{supertabular}
\usepackage{relsize}
\usepackage{psfrag}
\usepackage{color}
\usepackage{bbding}

\newcommand {\C} {{\rm I\kern-5.5pt C}}

\newcommand{\bcap} {\hspace{2pt} \mathlarger{\cap}
\hspace{2pt}}

\newcommand{\bP}[1]{{\mathbb{P}}\left[{#1}\right]}

\def\centerhack#1{\hbox to 0pt{\hss\footnotesize #1\hss}}
\def\centerhackn#1{\hbox to 0pt{\hss #1\hss}}
\def\dchack#1{\vbox to 0pt{\vss{\hbox to 0pt{\hss#1\hss}}\vss}}

\usepackage{enumitem}
\setlist{leftmargin=17pt}
\newcommand{\h}{it holds that }

\newtheorem{fact}{Fact}
\newtheorem{lem}{Lemma}
\newtheorem{thm}{Theorem}

\newtheorem{rem}{Remark}
\newtheorem{cor}{Corollary}
\newtheorem{proposition}{Proposition}
\newtheorem*{proposition1.1}{Proposition 1.1}
\newtheorem*{proposition1.2}{Proposition 1.2}
\newtheorem*{proposition1.3}{Proposition 1.3}
\newtheorem*{proposition2.1}{Proposition 2.1}
\newtheorem*{proposition2.2}{Proposition 2.2}
\begin{document}
%
\title{Topological Properties of Secure Wireless Sensor Networks under the
$q$-Composite Key Predistribution Scheme with Unreliable Links}
%
%
%

\author{Jun~Zhao,~\IEEEmembership{Member,~IEEE}
\thanks{Manuscript received March 16, 2016; revised October 24, 2016; accepted January 5, 2017; approved
by \textsc{\mbox{IEEE/ACM Transactions on Networking}} Editor Y. Yi. Date of
publication February 1, 2017; date of current version February 1, 2017. \newline \indent J. Zhao was with the Cybersecurity Lab (CyLab) at Carnegie Mellon University, Pittsburgh, PA 15213, USA. He is now with Arizona State University, Tempe, AZ 85281, USA (Email: junzhao@alumni.cmu.edu).
\newline \indent A preliminary version of this paper appeared as: \newline J.~Zhao, O.~Ya\u{g}an, and V.~Gligor, ``On Topological Properties of Wireless Sensor Networks under the
  $q$-Composite Key Predistribution Scheme with On/Off Channels,'' IEEE International Symposium on Information Theory, Hawaii, HI, USA, June 2014.\newline \indent This research was supported in part by Arizona State University, by   the U.S. National Science Foundation under Grant CNS-1422277, and by the U.S. Defense Threat
Reduction Agency under Grant HDTRA1-13-1-0029. This research was also supported in part by CyLab and Department of Electrical \& Computer Engineering
at Carnegie Mellon University.}}

%
%

\markboth{IEEE/ACM TRANSACTIONS ON NETWORKING}%
{ZHAO: Topological Properties of Secure Sensor Networks under
$q$-Composite Key Predistribution with Unreliable Links}
%



\maketitle


\begin{abstract}
Security is an important issue in wireless sensor networks (WSNs), which are often deployed in hostile environments. The $q$-composite key predistribution scheme has been recognized as a suitable approach to secure WSNs. Although
the $q$-composite scheme has received much attention in the literature, there is still a lack of rigorous analysis for secure
WSNs operating under the $q$-composite scheme in consideration of the unreliability of   links.  One main difficulty lies in analyzing the network topology whose links are not independent. Wireless links can be unreliable in practice due
to the presence of physical barriers between sensors or because
of harsh environmental conditions severely impairing communications. In this paper, we resolve the difficult challenge and
investigate topological properties related to node degree in WSNs
operating under the $q$-composite scheme with unreliable
communication links modeled as independent on/off channels.  Specifically, we derive the asymptotically exact probability for the property of minimum degree being at least $k$, present the asymptotic probability distribution for the minimum
degree, and
demonstrate that the number of nodes with an
arbitrary degree is in distribution  asymptotically equivalent to a
Poisson random variable. We further use the theoretical results to provide useful design guidelines for secure WSNs.  Experimental results also confirm the validity of our analytical findings.

\end{abstract}


%
\begin{IEEEkeywords}
Security, key predistribution,  wireless sensor networks, random graphs,
topological properties.
 \end{IEEEkeywords}

%
%
%
\section{Introduction}

Wireless sensor networks (WSNs) enable a broad range
of applications including military surveillance, home automation, and  patient monitoring \cite{adrian}. In many
scenarios, since WSNs are deployed in adversarial  environments, security becomes an important issue. To this end, key predistribution has been recognized as a typical solution
to secure WSNs \cite{virgil}.  The idea is to randomly assign cryptographic keys to sensors before
network deployment. Various key predistribution schemes have been studied in the literature \cite{ZhaoYaganGligor,Rybarczyk,yagan_onoff,zhao2016resilience,adrian,virgil,Krzywdzi,ryb3,zhao2015threshold,yagan,zhao2015resilience,nikoletseas2015some,ICASSP17-social}.

The $q$-composite key predistribution scheme proposed by Chan
\emph{et al.} \cite{adrian} as an extension of the Eschenauer-Gligor
scheme \cite{virgil} (the $q$-composite scheme in the case of $q=1$)
has received much interest \cite{yagan,zhao2015resilience,nikoletseas2015some,ICASSP17-social,bloznelis2013,GlobalSIP15-parameter,ISIT_RKGRGG,ICASSP17-design}
 since its introduction. The $q$-composite scheme when $q\geq 2$
outperforms the Eschenauer-Gligor scheme in terms of the strength
against small-scale network capture attacks while trading off
increased vulnerability in the face of large-scale attacks.

The $q$-composite scheme \cite{adrian} works as follows. For a WSN with $n$
sensors, prior to deployment, each sensor is independently assigned
$K_n$ different keys which are selected uniformly at random from a
pool of $P_n$ keys, where $K_n$ and $P_n$ are both functions of $n$,
with $K_n \leq P_n$. Then two sensors establish a link in between
after deployment if and only if they share at least $q$ keys
\emph{and} the physical link constraint between them is satisfied.
Examples of physical link constraints include the reliability of the
transmission channel \cite{ZhaoYaganGligor,yagan_onoff} and the requirement that the distance between two sensors should be close
enough for communication \cite{ISIT_RKGRGG}.

Communication links between sensor nodes may not be available due to the
presence of physical barriers between nodes or because of harsh
environmental conditions severely impairing transmission. To represent
unreliable links, we use the \emph{on}/\emph{off} channel model where each link is either {\em on} (i.e., {\em active}) with probability
$p_n$ or {\em off} (i.e., {\em inactive}) with probability $(1-p_n)$, where $p_n$ is a
function of $n$ for generality.

In addition to link failure, sensor nodes are also prone to failure in
 WSNs deployed
in hostile environments. To ensure reliability against the
failure of sensors, we study the property of minimum degree being at least $k$ so that each sensor is directly connected to at least $k$ other sensors. This means that a sensor may still be connected to a sufficient number of sensors even if some neighbors fail. Note that the degree of a node $v$ is the
number of nodes having links with $v$; and the minimum (node) degree
of a network is the least among the degrees of all nodes. Another related graph property is $k$-connectivity, which is stronger than the property of minimum degree being at least $k$.  A network (or
a graph) is said to be $k$-connected if it remains connected despite the deletion of any $(k - 1)$ nodes \cite{shahrivar2015robustness,dibaji2015consensus}; a network is simply deemed connected if it is $1$-connected. Hence, $k$-connectivity provides
a guarantee of network reliability against  the failure  of $(k - 1)$ sensors due to
adversarial attacks, battery depletion, harsh environmental conditions, etc.

In view of the above, we investigate topological properties related to node degree in WSNs employing the $q$-composite key predistribution scheme under the \emph{on}/\emph{off} channel model as the
physical link constraint comprising independent channels which are
either \emph{on} or \emph{off}.
Specifically,
we derive the asymptotically exact probabilities for the property of minimum degree being at least $k$, establish the asymptotic probability distribution for the minimum
degree, and
show that the number of nodes with an
arbitrary degree is in distribution  asymptotically equivalent to a
Poisson random variable. Our results are useful for designing secure WSNs under link and node failure.


We summarize our contributions in the following two subsections. We first present our results on node degree for a secure WSN employing the $q$-composite key predistribution scheme under the on/off channel model. Then we use the results to provide useful design guidelines for secure WSNs.

\subsection{\textbf{Results}}

For $\mathbb{G}_q$ denoting a secure sensor network with the $q$-composite key predistribution scheme under the on/off channel model, we present several results related to node degree, by considering the conditions on $p_{e, q}$, which denotes the probability of a secure link between two sensors.  The secure link probability $p_{e, q}$ is given by
$p_{e, q} = p_n \cdot \left[1- \sum_{u=0}^{q-1} \frac{\binom{K_n}{u}\binom{P_n-K_n}{K_n-u}}{\binom{P_n}{K_n}}\right]$,
as shown in Equation (\ref{psq2cijFC3}) on Page \pageref{psq2cijFC3} later.

For the network $\mathbb{G}_q$, we now present the results, which are further elaborated in Section \ref{sec-results-Gq}.
\begin{itemize}[leftmargin=7pt]
\item[$\bullet$] First, we derive the asymptotically exact probabilities for the property of minimum degree being at least $k$. Specifically,
if $p_{e, q}  = \frac{\ln  n + {(k-1)} \ln \ln n + {\alpha_n}}{n}$ for a constant integer $k \geq 1$ and a sequence $\alpha_n$ satisfying $\lim_{n \to \infty} \alpha_n  \in [-\infty, \infty]$, then the probability that  $\mathbb{G}_q$ has a minimum degree at least $k$ converges to $e^{- \frac{e^{-\lim_{n \to \infty} \alpha_n }}{(k-1)!}}$, which  equals (i) $e^{- \frac{e^{-\alpha ^*}}{(k-1)!}}$ if $\lim_{n \to \infty} \alpha_n = \alpha ^* \in (-\infty, \infty)$, (ii) $1$ if $\lim_{n \to \infty} \alpha_n = \infty$, and (iii) $0$ if $\lim_{n \to \infty} \alpha_n = -\infty$.  \vspace{1pt}
\item[$\bullet$] We  extend the above result to provide the asymptotic probability distribution for the minimum
degree. Specifically, when $\alpha_n$ above can be written as $\alpha_n = b \ln \ln n + \beta_n$ for
a constant integer $b$ and a sequence $\beta_n$
satisfying $-1     < \liminf_{n \to \infty} \frac{\beta_n}{\ln \ln n} \leq
\limsup_{n \to \infty} \frac{\beta_n}{\ln \ln n}< 1$ (i.e., $c_1 \ln \ln n \leq \beta_n \leq c_2 \ln \ln n $ for constants $-1<c_1 \leq c_2 < 1$), we have the following:
\begin{itemize}[leftmargin=9pt]
\item[$\mathsmaller{\bullet}$] if $k+b \geq 1$ and $\lim_{n \to \infty} \beta_n  \in [-\infty, \infty]$, then the minimum degree of $\mathbb{G}_q$ in the asymptotic sense equals (i) $k+b$ with probability $e^{- \frac{e^{-\lim_{n \to \infty} \beta_n }}{(k-1)!}}$, \vspace{1pt}  (ii) $k+b-1$ with probability $1-e^{- \frac{e^{-\lim_{n \to \infty} \beta_n }}{(k-1)!}}$, and (iii) other values with probability $0$;
\item[$\mathsmaller{\bullet}$] if $k+b \leq 0$, then the minimum degree of $\mathbb{G}_q$ in the asymptotic sense equals  $0$ with probability $1$. \vspace{1pt}
\end{itemize}
\item[$\bullet$] Our results on minimum degree are obtained by analyzing the number of nodes with a fixed degree. Specifically, we show that for a non-negative constant integer $h$, the number of nodes in
$\mathbb{G}_q$ with degree $h$ is in distribution asymptotically equivalent to a
Poisson random variable with mean $ n (h!)^{-1}(n p_{e, q})^h e^{-n
p_{e, q}}$.
\end{itemize}

\subsection{\textbf{Design guidelines for secure sensor networks}}

Based on the above results, for $\mathbb{G}_q$ denoting a secure sensor network employing the $q$-composite key predistribution scheme under the on/off channel model, we obtain several guidelines below for choosing parameters to ensure that the network $\mathbb{G}_q$ has certain minimum node degree. The guidelines are given by enforcing conditions on $p_{e, q}$, the probability of a secure link between two sensors.
Note that $p_{e, q} = p_n \cdot \left[1- \sum_{u=0}^{q-1} \frac{\binom{K_n}{u}\binom{P_n-K_n}{K_n-u}}{\binom{P_n}{K_n}}\right]$; see Equation (\ref{psq2cijFC3})  later.

For the network $\mathbb{G}_q$, we now present the design guidelines, which are further explained in Section \ref{sec-design-guidelines}.
\begin{itemize}[leftmargin=7pt]
\item[$\bullet$] First, to ensure that the network $\mathbb{G}_q$ has a minimum degree no less
than $k$ (i.e., to ensure that each sensor is directly connected to at least $k$ other sensors), we can choose network parameters to have
\begin{align}
p_{e, q} \geq \frac{\ln  n + {(k+c_1-1)} \ln \ln n  }{n} \text{ for a constant $c_1 > 0$,}
\label{peq1sbsc-critical1-repeat}
\end{align}
where the positive constant $c_1$ can be arbitrarily small. \vspace{4pt}
\item[$\bullet$] Second, to guarantee that the network $\mathbb{G}_q$ has a minimum degree at least $k$ with probability no less than $\rho$, we   choose   parameters to have
\begin{align}
p_{e, q} \geq \frac{\ln  n + {(k-1)} \ln \ln n - \ln [ (k-1)! \ln \frac{1}{\rho}] }{n}
\label{peq1sbsc-critical1-with-prob-rho-repeat}.
\end{align}
~\vspace{-6pt}
\item[$\bullet$] Third, to ensure that the network $\mathbb{G}_q$ has a minimum degree being $k$ \textit{exactly}, we can choose network parameters to have
\begin{align}
p_{e, q}  = \frac{\ln  n + {(k+c_2-1)} \ln \ln n  }{n}  \text{ for a constant $0<c_2 <1$,}
\label{peq1sbsc-critical2-repeat}
\end{align}
where the positive constant $c_2$ can be arbitrarily small.
\end{itemize}

\subsection{Roadmap}

We organize the rest of the paper as follows. Section
\ref{sec:SystemModel} describes the system model in detail.
Afterwards, we elaborate and discuss the results in Section
\ref{sec:res}. In Section \ref{sec-prove-first-two-theorems-mnd}, we prove Theorems \ref{thm:exact_qcomposite} and \ref{thm:exact_qcomposite-more-fine-grained} using Theorem \ref{thm:exact_qcomposite2}. In Section \ref{sec_est}, we detail the steps of
establishing Theorem \ref{thm:exact_qcomposite2} through Lemma \ref{LEM1}. Section
\ref{secprf:lem_pos_exp} provides the proof of Lemma \ref{LEM1} by
the help of Propositions \ref{PROP_ONE} and \ref{PROP_SND}, which are proved in Sections
\ref{sec:PROP_ONE} and \ref{sec:PROP_SND}, respectively.
Subsequently, we present experiments in Section
\ref{sec:expe} to confirm our analytical results. Section \ref{related} is devoted to relevant results
in the literature. Next, we conclude the paper and identify future
research directions in Section \ref{sec:Conclusion}, followed by
the Appendix.

\section{System Model}
\label{sec:SystemModel}

Our approach to the analysis is to explore the induced random graph
models of the WSNs. As will be clear soon, the graph modeling a WSN under $q$-composite
scheme and the on/off channel model is an intersection of two graphs
belonging to different kinds, which renders the analysis challenging
due to the intertwining of the two distinct types of random graphs
\cite{yagan_onoff,ISIT}.

We elaborate the graph modeling of a WSN with $n$ sensors, which
employs the $q$-composite key predistribution scheme and works under
the {on/off} channel model. We consider a node set $\mathcal {V} =
\{v_1, v_2, \ldots, v_n \}$ to represent the $n$ sensors (a sensor
is also referred to as a node). For each node $v_i \in \mathcal
{V}$, the set of its $K_n$ different keys is denoted by $S_i$, which
is uniformly distributed among all $K_n$-size subsets of a key pool
of $P_n$ keys, and is referred to as the key ring of node $v_i$.

The $q$-composite key predistribution scheme is
 modeled by a graph denoted by $G_q(n,K_n,P_n)$, which is defined on the
vertex set $\mathcal{V}$ such that any two different nodes $v_i$ and
$v_j$ sharing at least $q$ keys (such event is denoted by
$\Gamma_{ij}$) have an edge in between. With $S_{ij} : = S_{i} \cap
S_{j}$, event $\Gamma_{ij}$  equals $\big[ |S_{ij}| \geq q \big]$,
where $|A|$ with $A$ as a set means the cardinality of $A$.

As discussed, under the {on/off} channel model, each node-to-node channel
independently has probability $p_n $ of being {\em on} and
probability $(1-p_n)$ of being {\em off}, where $p_n$ is a function
of $n$. Denoting by ${B}_{i j}$ the event that the channel between
distinct nodes $v_i$ and $v_j$ is {\em on}, we have $\bP{C_{ij}} =
p_n$, where $\mathbb{P}[\mathcal {E}]$ denotes the probability that
event $\mathcal {E}$ happens, throughout the paper. The {on/off}
channel model is represented by an Erd\H{o}s-R\'enyi graph $G(n,
p_n)$ \cite{citeulike:4012374} defined on the node set $\mathcal{V}$
such that $v_i$ and $v_j$ have an edge in between if event $C_{ij}$
happens.

Finally, we denote by $\mathbb{G}_q(n, K_n, P_n,
p_n)$ the underlying graph of the $n$-node WSN operating under the
$q$-composite key predistribution scheme and the on/off channel
model. We often write $\mathbb{G}_q$ rather than $\mathbb{G}_q(n,
K_n, P_n, p_n)$ for notation brevity. Graph
$\mathbb{G}_q$ is defined on the node set
$\mathcal{V}$ such that there exists an edge between nodes $v_i$ and
$v_j$ if events $\Gamma_{ij}$ and $C_{ij}$ happen at the same time.
We set event $E_{ij} : = \Gamma_{ij} \cap C_{ij}$ and also write
$E_{ij} $ as $E_{v_i v_j} $ when necessary.
 It is clear that $\mathbb{G}_q$ can be seen as the
intersection of $G_q(n, K_n, P_n)$ and $G(n, p_n)$, meaning
\begin{equation}
\mathbb{G}_q= G_q(n, K_n, P_n) \cap G(n, p_n).
 \label{eq:G_on_is_RKG_cap_ER_oyton}
\end{equation}

We define $p_{s,q} $ as the probability that two different nodes
share at least $q$ keys and $p_{e,q} $ as the probability that two
distinct nodes have a link in between, where the subscripts ``s''
and ``e'' are short for ``secure'' and ``edge'', respectively.
 $p_{s,q} $ and $p_{e,q}$ both rely on $K_n, P_n$ and $q$, while
$p_{e,q}$ also depends on $p_n$. Under $P_n \geq 2 K_n$, we determine $p_{s,q}$ through
\begin{align}
p_{s,q}=  \mathbb{P} [\Gamma_{i j} ] &  = \sum_{u=q}^{K_n}
  \mathbb{P}[|S_{i} \cap S_{j}| = u] \nonumber \\ & = 1 - \sum_{u=0}^{q-1}
  \mathbb{P}[|S_{i} \cap S_{j}| = u] , \label{psq1}
\end{align}
where
\begin{align}
& \mathbb{P}[|S_{i} \cap S_{j}| = u] = \frac{\binom{K_n}{u}\binom{P_n-K_n}{K_n-u}}{\binom{P_n}{K_n}} ,
 \textrm{ for } u = 0,1, \ldots, K_n, \label{psq2}
\end{align}
since $S_{i}$ and $S_{j}$ are independently and uniformly selected
from all $K_n$-size subsets of a key pool with size $P_n$. Then by
the independence of events ${C}_{i j} $ and $ \Gamma_{i j} $, we
obtain
\begin{align}
{p_{e,q}}  & =  \mathbb{P} [E_{i j} ]  =  \mathbb{P} [{C}_{i j} ]
\cdot \mathbb{P} [\Gamma_{i j} ] =  p_n\cdot
p_{s,q}. \label{eq_pre}
\end{align}
Summarizing (\ref{psq1})  (\ref{psq2})  (\ref{eq_pre}), we derive that under $P_n \geq 2 K_n$, the link probability $p_{e,q}$ is given by
\begin{align}
 p_{e,q} &  =
 p_n \cdot \left[1- \sum_{u=0}^{q-1} \frac{\binom{K_n}{u}\binom{P_n-K_n}{K_n-u}}{\binom{P_n}{K_n}}\right]  .\label{psq2cijFC3}
 \end{align}

\section{The Results and Discussion} \label{sec:res}

We present and discuss the results in this section. Throughout the
paper, $q$ is a positive integer and does not scale with $n$;
$\mathbb{N}_0 $ stands for the set of all positive integers;
$\mathbb{R}$ is the set of all real numbers; $e$ is the base of the
natural logarithm function, $\ln$; and the floor function $\lfloor x
\rfloor$ is the largest integer not greater than $x$. We consider
$e^{\infty} = \infty$ and $e^{-\infty} = 0$. The term ``for all $n$
sufficiently large'' means ``for any $n \geq N$, where $N \in
\mathbb{N}_0$ is selected appropriately''. As already mentioned, all asymptotic statements are understood with $n \to \infty$, and we use the standard
asymptotic notation $o(\cdot), O(\cdot), \omega(\cdot),
\Omega(\cdot),\Theta(\cdot), \sim$; see \cite[Page 2-Footnote 1]{ZhaoYaganGligor}. In
particular, for two positive sequences $f_n$ and $g_n$, $f_n \sim
g_n$ signifies $\lim_{n \to
  \infty}\frac{{f_n}}{g_n}=1$; namely, $f_n$
  and $g_n$ are asymptotically equivalent.

\subsection{The Results of Graph
$\mathbb{G}_q\iffalse_{on}\fi$} \label{sec-results-Gq}

We now present the results of graph
$\mathbb{G}_q$ below.

Theorem \ref{thm:exact_qcomposite} provides the probability of minimum degree being at least $k$ in $\mathbb{G}_q$.

\begin{thm}[\textbf{Minimum degree in graph $\mathbb{G}_q\iffalse_{on}\fi$}]\label{thm:exact_qcomposite}
For graph $\mathbb{G}_q$ with $ K_n =
\omega(1)$ and $\frac{{K_n}^2}{P_n} = o(1)$,
if there exist a constant integer $k \geq 1$ and a sequence $\alpha_n$ satisfying $\lim\limits_{n \to \infty} \alpha_n  \in [-\infty, \infty]$ such that
\begin{align}
p_{e, q}  = \frac{\ln  n + {(k-1)} \ln \ln n + {\alpha_n}}{n},
\label{peq1sbsc}
\end{align}
 then with $\delta$ denoting the minimum degree of
$\mathbb{G}_q$, we have
\begin{align}
\hspace{-114pt}\lim\limits_{n \to \infty}\bP{\delta \geq  k}  \nonumber
 \end{align}
 ~\vspace{-10pt}
\begin{subnumcases}{\hspace{-14pt} = }
\hspace{-2pt}e^{- \frac{e^{-\alpha ^*}}{(k-1)!}}, &\hspace{-19pt} \textrm{if } $\lim\limits_{n \to \infty} \alpha_n\hspace{-1pt} =\hspace{-1pt} \alpha ^* \in (-\infty, \infty)$, \label{thm-mnd-alpha-finite} \\[2pt]
\hspace{-2pt}1, & \hspace{-19pt} \textrm{if } $\lim\limits_{n \to \infty} \alpha_n \hspace{-1pt}=\hspace{-1pt}  \infty$, \label{thm-mnd-alpha-infinite}\\[2pt]
\hspace{-2pt}0, & \hspace{-19pt} \textrm{if } $\lim\limits_{n \to \infty} \alpha_n \hspace{-1pt}= \hspace{-1pt}- \infty$. \label{thm-mnd-alpha-minus-infinite}
\end{subnumcases}
\end{thm}

\begin{rem} \label{thm:exact_qcomposite-rem}
The results (\ref{thm-mnd-alpha-finite}) (\ref{thm-mnd-alpha-infinite}) (\ref{thm-mnd-alpha-minus-infinite}) can be compactly summarized as $\lim_{n \to \infty}\bP{\delta \geq  k}=e^{- \frac{e^{-\lim_{n \to \infty} \alpha_n }}{(k-1)!}}$.
\end{rem}

\textbf{Interpreting Theorem \ref{thm:exact_qcomposite}.}
Theorem \ref{thm:exact_qcomposite} for graph
$\mathbb{G}_q$ presents the asymptotically exact
probability and a zero--one law for the event that
$\mathbb{G}_q$ has a minimum   degree no less
than $k$, where a zero--one law means that the probability of a graph having a certain property asymptotically converges to $0$ under some conditions and to $1$ under some other conditions.
To establish Theorem \ref{thm:exact_qcomposite}, we explain the basic ideas in Section \ref{sec-basic-proof-ideas}, and more technical details in Section \ref{sec-prove-first-two-theorems-mnd}.

While Theorem \ref{thm:exact_qcomposite} above is for the property of minimum degree being at least some value, we now present Theorem \ref{thm:exact_qcomposite-more-fine-grained} below, which gives a more fine-grained result to provide the asymptotic probability distribution for the minimum
degree.


\begin{thm}[\textbf{Minimum degree in graph $\mathbb{G}_q\iffalse_{on}\fi$: \textit{More fine-grained results} compared with Theorem \ref{thm:exact_qcomposite}}] \label{thm:exact_qcomposite-more-fine-grained}
Under the conditions of Theorem \ref{thm:exact_qcomposite}, if $\alpha_n$ in Equation (\ref{peq1sbsc}) can be written as
\begin{align}
\alpha_n = b \ln \ln n + \beta_n \label{alpha-n-written-beta-n}
\end{align}
for
a constant integer $b$ and a sequence $\beta_n$
satisfying
\begin{align}
-1     < \liminf_{n \to \infty} \frac{\beta_n}{\ln \ln n} \leq
\limsup_{n \to \infty} \frac{\beta_n}{\ln \ln n}< 1,  \label{liminfbetan}
\end{align}
then with $\delta$ denoting the minimum degree of
$\mathbb{G}_q$, the  properties \ding{172}--\ding{175} below follow:
\begin{itemize}
\item[\ding{172}] for ${k+b \leq 0}$ (which implies $b \leq -k \leq - 1$ given $k \geq 1$), we have
\begin{subnumcases}{}
\lim\limits_{n \to \infty}\bP{\delta =  0} =1, \label{thm-mnd-delta-0} \\
\lim\limits_{n \to \infty}\bP{\delta > 0}  =0; \label{thm-mnd-delta-non-0}
\end{subnumcases}
\item[]
\item[]\hspace{-20pt}and for ${k+b \geq 1}$  (i.e., $b \geq 1-k$), we obtain properties \ding{173}--\ding{176}:
\item[\ding{173}] $\lim\limits_{n \to \infty}\bP{(\delta =  k+b) \text{ or } (\delta =  k+b-1)}=1$;
\item[\ding{174}] if $\lim\limits_{n \to \infty} \beta_n = \beta ^* \in (-\infty, \infty)$, then
\begin{subnumcases}{}
\lim\limits_{n \to \infty}\bP{\delta =  k+b}  =e^{- \frac{e^{-\beta ^*}}{(k+b-1)!}}, \label{thm-mnd-beta-finite-1} \\
\lim\limits_{n \to \infty}\bP{\delta =  k+b-1}  =1-e^{- \frac{e^{-\beta ^*}}{(k+b-1)!}}; \label{thm-mnd-beta-finite-1}
\end{subnumcases}
\item[\ding{175}]
 if $ \lim\limits_{n \to \infty} \beta_n = \infty$, then
\begin{subnumcases}{}
\lim\limits_{n \to \infty}\bP{\delta =  k+b} =1, \label{thm-mnd-beta-infinite-1} \\
 \lim\limits_{n \to \infty}\bP{\delta  \neq  k+b }  =0; \label{thm-mnd-beta-infinite-2}
\end{subnumcases}
\item[\ding{176}] if $ \lim\limits_{n \to \infty} \beta_n = - \infty$, then
\begin{subnumcases}{}
\lim\limits_{n \to \infty}\bP{\delta =  k+b-1} =1, \label{thm-mnd-beta-minus-infinite-1} \\
 \lim\limits_{n \to \infty}\bP{\delta  \neq  k+b-1 }  =0.\label{thm-mnd-beta-minus-infinite-2}
\end{subnumcases}
\end{itemize}
\end{thm}

\begin{rem} \label{thm:exact_qcomposite-more-fine-grained-rem}
 The above results \ding{174}--\ding{176} for $k+b \geq 1$ and $\lim_{n \to \infty} \beta_n  \in [-\infty, \infty]$ can be compactly summarized as that the minimum degree of $\mathbb{G}_q$ in the asymptotic sense equals \vspace{1pt}  (i) $k+b$ with probability $e^{- \frac{e^{-\lim_{n \to \infty} \beta_n }}{(k-1)!}}$, (ii) $k+b-1$ with probability $1-e^{- \frac{e^{-\lim_{n \to \infty} \beta_n }}{(k-1)!}}$, and (iii) other values with probability $0$, while results \ding{172} says that if $k+b \leq 0$, then the minimum degree of $\mathbb{G}_q$ in the asymptotic sense equals  $0$ with probability $1$.
\end{rem}

\textbf{Interpreting Theorem \ref{thm:exact_qcomposite-more-fine-grained}.} Theorem \ref{thm:exact_qcomposite-more-fine-grained} presents the asymptotic probability distribution for the minimum
degree. We   explain that Theorem \ref{thm:exact_qcomposite-more-fine-grained} is more fine-grained than Theorem~\ref{thm:exact_qcomposite}. We discuss first Theorem \ref{thm:exact_qcomposite-more-fine-grained}'s result \ding{172} and then its results \ding{173}--\ding{176}.
\begin{itemize}
\item[\textbf{(i)}] In result \ding{172} above, $b \leq -k \leq - 1$ follows from $k+b \leq 0$ and $k \geq 1$. Using $b \leq - 1$ and (\ref{liminfbetan}) in  (\ref{alpha-n-written-beta-n}), we have $\lim_{n \to \infty} \alpha_n = - \infty$, so we use (\ref{thm-mnd-alpha-minus-infinite}) of Theorem \ref{thm:exact_qcomposite} to obtain $\delta<k$ almost surely (an event happens \textit{almost surely} if its probability converges $1$ as $n \to \infty$), where $\delta$ denotes the minimum degree of
$\mathbb{G}_q$. For comparison, (\ref{thm-mnd-delta-0}) of Theorem \ref{thm:exact_qcomposite-more-fine-grained} presents the stronger result that $\delta=0$ almost surely.
\item[\textbf{(ii)}] In the above results \ding{173}--\ding{176} where ${k+b \geq 1}$ holds (i.e., $b \geq 1-k$) , we derive from (\ref{alpha-n-written-beta-n}) and (\ref{liminfbetan}) that
\begin{subnumcases}{\hspace{-37pt}\lim\limits_{n \to \infty} \alpha_n\hspace{-2pt} =\hspace{-2pt}}\hspace{-2pt} \infty,
&\hspace{-19pt}\text{if $b\hspace{-2pt}\geq \hspace{-2pt} 1$,} \label{b-geq-1}  \\
\hspace{-2pt}\beta ^* ,
&\hspace{-19pt}\text{if $b\hspace{-2pt}=\hspace{-2pt}0$ and $ \lim\limits_{n \to \infty} \beta_n\hspace{-2pt} = \hspace{-2pt}\beta ^* \hspace{-2pt}\in \hspace{-2pt}(-\infty, \infty)$,} \label{b-eq-0-beta-finite}  \\
\hspace{-2pt}\infty,
&\hspace{-19pt}\text{if $b\hspace{-2pt}=\hspace{-2pt}0$ and $ \lim\limits_{n \to \infty} \beta_n\hspace{-2pt} =\hspace{-2pt} \infty$,} \label{b-eq-0-beta-infinite}  \\
\hspace{-2pt}-\infty,
&\hspace{-19pt}\text{if $b\hspace{-2pt}=\hspace{-2pt}0$ and $ \lim\limits_{n \to \infty} \beta_n \hspace{-2pt}=\hspace{-2pt} -\infty$,} \label{b-eq-0-beta-minus-infinite}  \\ \hspace{-2pt}-\infty,
&\hspace{-19pt}\text{if $ b\hspace{-2pt}\leq \hspace{-2pt}-1$.} \label{b-leq-minus-1}  \end{subnumcases}
Below we discuss (\ref{b-geq-1})--(\ref{b-leq-minus-1}), respectively.
\begin{itemize}[leftmargin=0pt]
\item[$\bullet$] For {(\ref{b-geq-1})} above, (\ref{thm-mnd-alpha-infinite}) of Theorem \ref{thm:exact_qcomposite} says $\delta\geq k$ almost surely, while \ding{173} of Theorem \ref{thm:exact_qcomposite-more-fine-grained} presents the stronger result that $\delta$ equals $k+b$ or $k+b-1$ almost surely (note $b\geq 1$ in (\ref{b-geq-1})).
\item[$\bullet$] For {(\ref{b-eq-0-beta-finite})} above, (\ref{thm-mnd-alpha-finite}) of Theorem \ref{thm:exact_qcomposite} says $\delta\geq k$  with probability $e^{- \frac{e^{-\beta ^*}}{(k+b-1)!}}$ asymptotically, while \ding{174} of Theorem \ref{thm:exact_qcomposite-more-fine-grained} presents the stronger result that $\delta$ equals $k$ (note $b=0$ in (\ref{b-eq-0-beta-finite}) here) with probability $e^{- \frac{e^{-\beta ^*}}{(k+b-1)!}}$ asymptotically, and equals $k-1$ with probability $1-e^{- \frac{e^{-\beta ^*}}{(k+b-1)!}}$ asymptotically (note $b=0$ in (\ref{b-eq-0-beta-finite}) here).
\item[$\bullet$] For {(\ref{b-eq-0-beta-infinite})} above, (\ref{thm-mnd-alpha-infinite}) of Theorem \ref{thm:exact_qcomposite} says $\delta\geq k$  almost surely, while \ding{175} of Theorem \ref{thm:exact_qcomposite-more-fine-grained} presents the stronger result that $\delta=k$ almost surely (note $b=0$ in (\ref{b-eq-0-beta-infinite}) here).
\item[$\bullet$] For {(\ref{b-eq-0-beta-minus-infinite})} above, (\ref{thm-mnd-alpha-minus-infinite}) of Theorem \ref{thm:exact_qcomposite} says $\delta < k$  almost surely, while \ding{176} of Theorem \ref{thm:exact_qcomposite-more-fine-grained} presents the stronger result that $\delta=k-1$ almost surely (note $b=0$ in (\ref{b-eq-0-beta-minus-infinite}) here).
\item[$\bullet$] For {(\ref{b-leq-minus-1})} above, (\ref{thm-mnd-alpha-minus-infinite}) of Theorem \ref{thm:exact_qcomposite} says $\delta < k$  almost surely, while \ding{173} of Theorem \ref{thm:exact_qcomposite-more-fine-grained}  presents the stronger result that $\delta$ equals $k+b$ or $k+b-1$ almost surely (note $b\leq -1$ in (\ref{b-leq-minus-1})).
\end{itemize}
\end{itemize}
Summarizing the above, compared with Theorem \ref{thm:exact_qcomposite}, Theorem~\ref{thm:exact_qcomposite-more-fine-grained} presents a more fine-grained result for
minimum degree in $\mathbb{G}_q$.

%
%
%
%

To prove Theorem \ref{thm:exact_qcomposite-more-fine-grained}, we provide the basic ideas in Section \ref{sec-basic-proof-ideas}, and more technical details in Section \ref{sec-prove-first-two-theorems-mnd}.

Theorems \ref{thm:exact_qcomposite} and \ref{thm:exact_qcomposite-more-fine-grained} above are for the property of minimum degree being at least $k$. We now consider a stronger graph/network property, namely $k$-connectivity.

\textbf{Extension to $k$-connectivity.} We can extend Theorem  \ref{thm:exact_qcomposite} to obtain the probability of $k$-connectivity in $\mathbb{G}_q$. Specifically, we can replace $\lim_{n \to \infty}\bP{\delta \geq  k}$ by $\lim_{n \to \infty}\bP{\text{$\mathbb{G}_q\iffalse_{on}^{(q)}\fi$  is $k$-connected.}}$, \vspace{2pt} at the cost of replacing $ K_n =
\omega(1)$ and $\frac{{K_n}^2}{P_n} = o(1)$ by a stronger condition set $ \frac{{K_n}^2}{P_n}  =
 o\left( \frac{1}{\ln n} \right)$, $ \frac{K_n}{P_n} = o\left( \frac{1}{n\ln n} \right)$ and $ K_n =
\Omega(n^{\epsilon})$ \vspace{2pt} for a positive constant $\epsilon$. Due to space limitation, we present the proof in the full version \cite{full}.

\subsection{{Design guidelines for secure sensor networks}} \label{sec-design-guidelines}

Based on the above results, now we provide several design guidelines of
secure sensor networks for achieving certain strength of minimum degree.

\begin{itemize}[leftmargin=2pt]
\item First, to ensure that $\mathbb{G}_q$ has a minimum degree no less
than $k$, we can choose network parameters to set
\begin{align}
p_{e, q} \geq \frac{\ln  n + {(k+c_1-1)} \ln \ln n  }{n} \text{ for a constant $c_1 > 0$,}
\label{peq1sbsc-critical1}
\end{align}
where the positive constant $c_1$ can be arbitrarily small. To see this, since (\ref{peq1sbsc-critical1}) implies that
$\alpha_n$ defined by (\ref{peq1sbsc}) (i.e., $p_{e, q}  = \frac{\ln  n + {(k-1)} \ln \ln n + {\alpha_n}}{n}$) satisfies $\lim_{n \to \infty} \alpha_n  =  \infty$, we use Theorem \ref{thm:exact_qcomposite} to have \vspace{4pt} $\bP{\text{$\mathbb{G}_q\iffalse_{on}^{(q)}\fi$  has a minimum degree at least $k$.}}=1$.
\item Second, to guarantee that $\mathbb{G}_q$ has a minimum degree at least $k$ with probability no less than $\rho$, we   choose   parameters to ensure
\begin{align}
p_{e, q} \geq \frac{\ln  n + {(k-1)} \ln \ln n - \ln [ (k-1)! \ln \frac{1}{\rho}] }{n}
\label{peq1sbsc-critical1-with-prob-rho}.
\end{align}
To see this, since (\ref{peq1sbsc-critical1-with-prob-rho}) implies that
$\alpha_n$ defined by (\ref{peq1sbsc}) (i.e., $p_{e, q}  = \frac{\ln  n + {(k-1)} \ln \ln n + {\alpha_n}}{n}$) satisfies $\alpha_n  \geq - \ln [ (k-1)! \ln \frac{1}{\rho}] $, we use Theorem \ref{thm:exact_qcomposite} to obtain \vspace{2pt} $\bP{\text{$\mathbb{G}_q\iffalse_{on}^{(q)}\fi$  has a minimum degree at least $k$.}} \geq e^{- \frac{e^{\ln [ (k-1)! \ln \frac{1}{\rho}]}}{(k-1)!}} = \rho$.\vspace{4pt}
\item Third, to ensure that $\mathbb{G}_q$ has a minimum degree being $k$ \textit{exactly}, we can choose network parameters to have
\begin{align}
p_{e, q}  = \frac{\ln  n + {(k+c_2-1)} \ln \ln n  }{n}  \text{ for a constant $0<c_2 <1$,}
\label{peq1sbsc-critical2}
\end{align}
where the positive constant $c_2$ can be arbitrarily small. To see this, (\ref{peq1sbsc-critical2}) implies \vspace{2pt} that
$\alpha_n$ defined by (\ref{peq1sbsc}) (i.e., $p_{e, q}  = \frac{\ln  n + {(k-1)} \ln \ln n + {\alpha_n}}{n}$) equals $c_2 \ln \ln n $, so $b$ in (\ref{alpha-n-written-beta-n}) is $0$ with $\beta_n$ satisfying (\ref{alpha-n-written-beta-n}) and $ \lim_{n \to \infty} \beta_n = \infty$. Then we use Theorem \ref{thm:exact_qcomposite}-Result \ding{175} to obtain $\bP{\text{Minimum degree of $\mathbb{G}_q\iffalse_{on}^{(q)}\fi$ equals  $k$ \textit{exactly}.}}=1$.
\end{itemize}


\subsection{{Basic Ideas to Establish Theorems \ref{thm:exact_qcomposite} and \ref{thm:exact_qcomposite-more-fine-grained}}}  \label{sec-basic-proof-ideas}


 We establish Theorems \ref{thm:exact_qcomposite} and \ref{thm:exact_qcomposite-more-fine-grained} for minimum degree in graph $\mathbb{G}_q$ by analyzing the number of nodes with a fixed degree, for which we present Theorem \ref{thm:exact_qcomposite2} below. The details of using Theorem \ref{thm:exact_qcomposite2} to prove Theorems \ref{thm:exact_qcomposite} and \ref{thm:exact_qcomposite-more-fine-grained} are given in Section \ref{sec-prove-first-two-theorems-mnd}.

\begin{thm}[\textbf{Possion distribution for number of nodes with a fixed degree in graph $\mathbb{G}_q\iffalse_{on}^{(q)}\fi$}]  \label{thm:exact_qcomposite2}
For graph $\mathbb{G}_q$ with $ K_n =
\omega(1)$  and $\frac{{K_n}^2}{P_n} = o(1)$, if
\begin{align}
p_{e, q} & = \frac{\ln  n \pm O(\ln \ln n)}{n} \label{peq1}
\end{align}
(i.e., $\frac{n p_{e, q} - \ln n }{\ln \ln n}$ is bounded), then for a non-negative constant integer $h$, the number of nodes in
$\mathbb{G}_q$ with degree $h$ is in distribution asymptotically equivalent to a
Poisson random variable with mean $\lambda_{n,h} : = n (h!)^{-1}(n p_{e, q})^h e^{-n
p_{e, q}}$; i.e., as $n \to \infty$,
\begin{align}
& \bP{\hspace{-3pt}\begin{array}{l}\text{The number of nodes in
$\mathbb{G}_q\iffalse_{on}\fi$}\\\text{with degree $h$ equals $\ell$.}\end{array}\hspace{-3pt}}\Big/\Big[(\ell !)^{-1}{\lambda_{n,h}} ^{\ell}e^{-\lambda_{n,h}}\Big] \to 1, \nonumber \\ & \text{~~~~~~~~~~~~~~~~~~~~~~~~~~~~~~~~~~~~~~~~~~~for $\ell = 0,1, \ldots$}  \label{eq-Poisson-lemma}
\end{align}
\end{thm}

\textbf{Interpreting Theorem \ref{thm:exact_qcomposite2}.}
Theorem \ref{thm:exact_qcomposite2} for graph
$\mathbb{G}_q$ shows that the number of nodes
with a fixed degree follows a Poisson distribution
asymptotically.

\subsection{The Practicality of the Theorem Conditions} \label{sec-Conditions-Practicality}

We check the practicality of the conditions in
Theorem \ref{thm:exact_qcomposite}: $ K_n =
\omega(1)$ and $\frac{{K_n}^2}{P_n} = o(1)$. The condition $ K_n = \omega(1)$ means that the key ring size $K_n$ on a sensor grows with the number  $n$ of sensors and thus it follows
trivially in secure wireless sensor networks \cite{DiPietro:2008:RSN:1341731.1341734,yagan,zhao2016topology}. For $k$-connectivity, the condition on $K_n$ (i.e., $ K_n =
\Omega(n^{\epsilon})$ is less appealing but is not much a problem because $\epsilon$ can be arbitrarily small.
In addition,  $\frac{{K_n}^2}{P_n}=
 o\left( \frac{1}{\ln n} \right)=o(1)$ and $ \frac{K_n}{P_n} = o\left( \frac{1}{n\ln n} \right)$
hold in practice since the key pool size $P_n$ is expected to
be several orders of magnitude larger than the key ring size $K_n$ (see \cite[Section 2.1]{virgil} and \cite[Section III-B]{yagan_onoff}).

\section{Proofs of Theorems \ref{thm:exact_qcomposite} and \ref{thm:exact_qcomposite-more-fine-grained} Using Theorem \ref{thm:exact_qcomposite2}} \label{sec-prove-first-two-theorems-mnd}

As explained in Section \ref{sec-basic-proof-ideas}, we establish Theorems \ref{thm:exact_qcomposite} and \ref{thm:exact_qcomposite-more-fine-grained} based on Theorem \ref{thm:exact_qcomposite2}.
Theorems \ref{thm:exact_qcomposite} and \ref{thm:exact_qcomposite-more-fine-grained} present results of $\delta$, where $\delta$ denotes the minimum degree of
$\mathbb{G}_q$. With $\Phi_{n,h}$ denoting the number of nodes with degree $h$ in $\mathbb{G}_{q}$, Theorem \ref{thm:exact_qcomposite2} provides the asymptotic distribution of $\Phi_{n,h}$. To use Theorem \ref{thm:exact_qcomposite2} for proving Theorems \ref{thm:exact_qcomposite} and \ref{thm:exact_qcomposite-more-fine-grained}, we now discuss the relationship between $\delta$ and $\Phi_{n,h}$. For non-negative integer $\mu$, it is straightforward to see properties \ding{202} and \ding{203} below.
\begin{itemize}[leftmargin=8pt]
  \item [\ding {202}]  The event $(\delta \geq \mu)$ (i.e., the event that the
minimum node degree of graph $\mathbb{G}_q$ is at
least $\mu$) is equivalent to the event $
\bigcap_{h=0}^{\mu-1} (\Phi_{n,h} = 0) $ (i.e., no node has degree
falling in $\{0,1,\ldots, \mu-1\}$).
  \item [\ding {203}] The
event $(\delta \leq \mu)$ (i.e., the event that the minimum node
degree of graph $\mathbb{G}_q$ is at most $\mu$)
and the event $ \bigcup_{h=0}^{\mu} (\Phi_{n,h} \neq 0) $ (i.e.,
there is at least one node with degree at most $\mu$) are
equivalent.
\end{itemize}

 Therefore, for any integer $\xi$, we obtain
\begin{align}
 &\mathbb{P}[\delta \geq \xi +1]  \nonumber  \\&=
\mathbb{P}\bigg[\bigcap_{h=0}^{\xi } (\Phi_{n,h} = 0)\bigg]
\textrm{~(by property \ding {202})} \nonumber
\\&   \leq  \mathbb{P}[\Phi_{n,\xi} = 0],\textrm{ if }\xi \geq 0 ,
\label{eqpd1} \end{align}\begin{align}
 & \mathbb{P}[\delta \leq \xi -2]  \nonumber  \\&\leq  \mathbb{P}\bigg[
\bigcup_{h=0}^{\xi-2} (\Phi_{n,h} \neq 0)\bigg] \textrm{~(by property
\ding {203})} \nonumber  \\&  \leq \sum_{h=0}^{\xi-2} \mathbb{P}[
\Phi_{n,h} \neq 0]\, \textrm{~(by the union bound)},\textrm{ if }\xi \geq 2,   \label{eqpd2}\end{align}\begin{align}
 & \mathbb{P}[\delta \geq \xi]\nonumber  \\& =
\mathbb{P}\bigg[\bigcap_{h=0}^{\xi-1} (\Phi_{n,h} = 0)\bigg]
\textrm{~(by property \ding {202})} \nonumber
\\&   \leq   \mathbb{P}[\Phi_{n,\xi-1} = 0],\textrm{ if }\xi \geq 1,
\label{eqn_1mindel2}
 \end{align}
and
\begin{align}
\mathbb{P}[\delta \geq \xi] & =
\mathbb{P}\bigg[\bigcap_{h=0}^{\xi-1}
(\Phi_{n,h} = 0)\bigg]  \textrm{~(by property \ding {202})}  \nonumber  \\
 & = 1 - \mathbb{P}\bigg[\bigcup_{h=0}^{\xi-1}
(\Phi_{n,h} \neq 0)\bigg] \nonumber  \\
 & \geq 1 - \sum_{h=0}^{\xi-1} \mathbb{P}[\Phi_{n,h} \neq 0]\textrm{ (by the union bound)} \nonumber
  \\  & =
 \mathbb{P}[\Phi_{n,k-1} = 0] - \boldsymbol{1}[k\geq 2]\times  \sum_{h=0}^{k-2}
\mathbb{P}[  \Phi_{n,h} \neq 0] ,  \label{eqn_1min}
\end{align}
where the indicator variable $\boldsymbol{1}[k\geq 2]$ equals $1$ if $k \geq 2$ and $0$ if $k<2$.

To use (\ref{eqpd1})--(\ref{eqn_1min}), we will compute $\mathbb{P}[
\Phi_{n,h} = 0]$ and $\mathbb{P}[
\Phi_{n,h} \neq 0]$ for $h=0,1,\ldots$ To this end, we use
Theorem \ref{thm:exact_qcomposite2}, which shows that $\Phi_{n,h}$ is in distribution asymptotically equivalent to a
Poisson random variable with mean $\lambda_{n,h}  $ specified by
 \begin{align}
 \lambda_{n,h} & : = n (h!)^{-1}(n p_{e, q})^h e^{-n
p_{e, q}}; \label{eqn_labmdahnew}
 \end{align}
 i.e.,
\begin{align}
 \mathbb{P}[\Phi_{n,h} = \ell]
 & \sim (\ell !)^{-1}{\lambda_{n,h}} ^{\ell} e^{-\lambda_{n,h}} .
 \label{eqn_phihellnew}
 \end{align}

To  assess $\lambda_{n,h}$ in (\ref{eqn_labmdahnew}), we use
(\ref{peq1sbsc}) about $p_{e, q}$ (i.e., $p_{e, q}   = \frac{\ln  n + {(k-1)} \ln \ln n + {\alpha_n}}{n}$). While $\alpha_n$ in Theorem \ref{thm:exact_qcomposite-more-fine-grained}  is given by (\ref{alpha-n-written-beta-n}) and satisfies $ |\alpha_n| = O(\ln \ln n) = o(\ln n)$ for
a constant integer $b$ and a sequence $\beta_n$ under
(\ref{liminfbetan}), $\alpha_n$ in Theorem \ref{thm:exact_qcomposite} may not satisfy  $|\alpha_n| = o(\ln n)$. However, we can still introduce the additional condition $|\alpha_n| = o(\ln n)$ in proving Theorem \ref{thm:exact_qcomposite}, as explained in Appendix E of the full version \cite{full}. The idea is to show that whenever Theorem \ref{thm:exact_qcomposite} with $|\alpha_n| = o(\ln n)$ holds, then Theorem \ref{thm:exact_qcomposite} regardless of $|\alpha_n| = o(\ln n)$. Now under $|\alpha_n| = o(\ln n)$ in Theorem \ref{thm:exact_qcomposite}, we   obtain
\begin{align}
p_{e, q} & \sim  \frac{\ln n}{n},\label{eq_pe_lnnn-tonnew}
\end{align}
where $f_n \sim
g_n$ for two positive sequences $f_n$ and $g_n$ means $\lim_{n \to
  \infty} {{f_n}}/{g_n}=1$; i.e., (\ref{eq_pe_lnnn-tonnew}) means
$\lim_{n \to
  \infty} {p_{e, q}}\big/\big(\frac{\ln n}{n}\big)=1$.

Then we substitute
(\ref{peq1sbsc}) and (\ref{eq_pe_lnnn-tonnew}) into (\ref{eqn_labmdahnew}) to derive
\begin{align}
 \lambda_{n,h}
 &  = n (h!)^{-1}(n p_{e,q})^h
e^{-n p_{e,q}} \nonumber  \\
 &  \sim n (h!)^{-1} (\ln n)^h \times e^{-\ln n -
 (k-1)\ln \ln n - \alpha_n}  \nonumber  \\
 & = (h!)^{-1} (\ln n)^{h+1-k} e^{-\alpha_n}. \label{liminfbetan3}
\end{align}


We now {use Theorem \ref{thm:exact_qcomposite2} (i.e., (\ref{eqn_phihellnew})) to prove Theorem \ref{thm:exact_qcomposite}} under the additional condition $|\alpha_n| = o(\ln n)$, which we can introduce based on the  above discussion. Then we evaluate $\mathbb{P}[\delta \geq k  ].$~Given $k \geq 1$, we know from (\ref{eqn_1mindel2}) and (\ref{eqn_phihellnew}) that
 \begin{align}
\mathbb{P}[\delta \geq k  ] &  \leq   e^{-\lambda_{n,k-1}} \times [1+o(1)],  \label{eqn_phihell-ton-part1}
\end{align}
and
know from (\ref{eqn_1min}) and (\ref{eqn_phihellnew}) that
 \begin{align}
 &\hspace{-8pt}  \mathbb{P}[\delta \geq k  ] \nonumber \geq \\  & \hspace{-8pt}  e^{-\lambda_{n,k-1}} \hspace{-2pt}\times \hspace{-2pt}[1\hspace{-2pt}-\hspace{-2pt}o(1)] \hspace{-1pt}- \hspace{-1pt}\boldsymbol{1}[ k\hspace{-2pt}\geq \hspace{-2pt}2]\hspace{-2pt}\times \hspace{-2pt} \sum_{h=0}^{k-2}
\big\{ \big(1- e^{-\lambda_{n,h}} \big)\hspace{-2pt} \times\hspace{-2pt} [1\hspace{-2pt}+\hspace{-2pt}o(1)]\big\} .   \label{eqn_phihell-ton-part2}
\end{align}
Based on (\ref{eqn_phihell-ton-part1}) and (\ref{eqn_phihell-ton-part2}), we discuss the following cases.
\begin{itemize}
\item If $\lim_{n \to \infty} \alpha_n  = \alpha ^* \in (-\infty, \infty)$,   (\ref{liminfbetan3}) implies for $k\geq 1$ that
\begin{subnumcases}{\hspace{-30pt}\lambda_{n,h} \to }
\hspace{-3pt} 0,&\hspace{-15pt}\textrm{for }$h = 0, 1,
\ldots,  k-2,$\textrm{ if $k \geq 2$}, \label{eqn-prove-thm1-alpha-n-finite-1} \\
\hspace{-3pt} \frac{e^{-\alpha ^*}}{(k-1)!},&\hspace{-15pt}\textrm{for $h = k-1$}.\label{eqn-prove-thm1-alpha-n-finite-2}
\end{subnumcases}
Applying (\ref{eqn-prove-thm1-alpha-n-finite-2}) to (\ref{eqn_phihell-ton-part1}), and applying (\ref{eqn-prove-thm1-alpha-n-finite-1}) (\ref{eqn-prove-thm1-alpha-n-finite-2}) to (\ref{eqn_phihell-ton-part2}), we have $e^{-
\frac{e^{-\alpha ^*}}{(k-1)!}} \times [1-o(1)] \leq \mathbb{P}[\delta \geq k] \leq e^{-
\frac{e^{-\alpha ^*}}{(k-1)!}} \times [1+o(1)]$ so that $\lim_{n \to \infty} \mathbb{P}[\delta \geq k]  = e^{-
\frac{e^{-\alpha ^*}}{(k-1)!}}$; i.e., (\ref{thm-mnd-alpha-finite}) is proved.
\item If $\lim_{n \to \infty} \alpha_n  = \infty$, then (\ref{liminfbetan3}) implies for $k\geq 1$ that
\begin{align}
\lambda_{n,h} \to  0 \textrm{~~~for } h = 0, 1,
\ldots,  k-1. \label{eqn-prove-thm1-alpha-n-infinite}
\end{align}
Substituting (\ref{eqn-prove-thm1-alpha-n-infinite}) into (\ref{eqn_phihell-ton-part1}), and substituting (\ref{eqn-prove-thm1-alpha-n-infinite}) into (\ref{eqn_phihell-ton-part2}), we obtain $  [1-o(1)] \leq \mathbb{P}[\delta \geq k] \leq  [1+o(1)]$ so that $\lim_{n \to \infty} \mathbb{P}[\delta \geq k]  = 1$; i.e., (\ref{thm-mnd-alpha-infinite}) is proved.
\item If $\lim_{n \to \infty} \alpha_n  = - \infty$, then (\ref{liminfbetan3}) implies for $k\geq 1$ that
\begin{align}
\lambda_{n,k-1} \to  \infty. \label{eqn-prove-thm1-alpha-n-minus-infinite}
\end{align}
Using (\ref{eqn-prove-thm1-alpha-n-minus-infinite} ) in (\ref{eqn_phihell-ton-part1}), we have $ \mathbb{P}[\delta \geq k] \leq o(1)$ so that $\lim_{n \to \infty} \mathbb{P}[\delta \geq k]  = 0$; i.e., (\ref{thm-mnd-alpha-minus-infinite}) is proved.
\item[]
\end{itemize}


 We now {use Theorem \ref{thm:exact_qcomposite2} (i.e., (\ref{eqn_phihellnew})) to prove Theorem \ref{thm:exact_qcomposite-more-fine-grained}}.
The condition (\ref{liminfbetan}) on $\beta_n$ implies that there are constants $c_1$ and $c_2$ with $-1<c_1\leq c_2 <1$ such that
\begin{align}
 c_1 \ln \ln n & \leq \beta_n \leq c_2 \ln \ln n, \text{ for all $n$ sufficiently large},  \label{liminfbetan2-pre}
\end{align}
which implies
\begin{align}
\hspace{-10pt} (\ln n)^{-c_2} & \leq e^{-\beta_n} \leq  (\ln n)^{-c_1}, \text{ for all $n$ sufficiently large}.  \label{liminfbetan2}
\end{align}
Using (\ref{alpha-n-written-beta-n}) in (\ref{peq1sbsc}), we have $p_{e, q}  = \frac{\ln  n + {(k+b-1)} \ln \ln n + {\beta_n}}{n}$, which along with (\ref{liminfbetan2-pre}) implies $p_{e, q} \sim \frac{\ln  n}{n}$. \vspace{1pt} Then similar to (\ref{liminfbetan3}), we derive $ \lambda_{n,h}\sim (h!)^{-1} (\ln n)^{h+1-(k+b)} e^{-\beta_n}$.
Applying (\ref{liminfbetan2}) to this result and noting $-1<c_1\leq c_2 <1$, we find
  \begin{align}
\hspace{-9pt} \lambda_{n,h} & \begin{cases} \to 0,&\hspace{-22pt}\textrm{for }h = 0, 1,
\ldots,  k+b-2,\textrm{ if }k+b \geq 2;  \\
\sim \frac{e^{-\beta_n}}{(k+b-1)!},&\hspace{-6pt}\textrm{for }h = k+b-1,\textrm{ if }k+b \geq 1;
 \\ \to \infty ,&\hspace{-22pt}\textrm{for }h = \max\{ k+b, 0\}, \max\{ k+b, 0\}+1, \ldots.
\end{cases} \label{eqn_lbdh}
 \end{align}
Using (\ref{eqn_lbdh}) in (\ref{eqn_phihellnew}), we get
  \begin{align}
\hspace{-129pt}  \mathbb{P}[\Phi_{n,h} = 0]  \sim e^{-\lambda_{n,h}} \nonumber
 \end{align}
\begin{subnumcases}{\hspace{-19pt}}
\hspace{-4pt} \to  1,\hspace{-32pt} \quad\quad\quad \textrm{ for }h = 0, 1,
\ldots,  k+b-2,\textrm{ if }k+b \geq 2; \label{eqn_expr_lahkk1case1}\\
\hspace{-4pt}\sim e^{-\frac{e^{-\beta_n}}{(k+b-1)!}}, \hspace{-1pt}\quad\textrm{for }h = k+b-1,\textrm{ if }k+b \geq 1; \label{eqn_expr_lahkk1case2}
 \\ \hspace{-4pt}\to 0 ,\hspace{-32pt} \quad\quad\quad\textrm{ for }h = \max\{ k+b, 0\}, \max\{ k+b, 0\}+1,
\ldots \label{eqn_expr_lahkk1case3}
\end{subnumcases}

If $k+b \leq 0$, then (\ref{eqn_expr_lahkk1case3}) gives $ \mathbb{P}[\Phi_{n,0} = 0] \to 0$, which along with (\ref{eqn_1mindel2}) and (\ref{eqn_1min}) yields $ \mathbb{P}[\delta \geq 1] = \mathbb{P}[\Phi_{n,0} = 0] \to 0$ so that we further obtain $\lim\limits_{n \to \infty}\bP{\delta =  0}=1$ and $\lim\limits_{n \to \infty}\bP{\delta > 0}=0$. Hence, property \ding{172} of Theorem \ref{thm:exact_qcomposite-more-fine-grained} is proved.

Below we consider the case of $k+b \geq 1$ to prove properties \ding{173}--\ding{176} of Theorem \ref{thm:exact_qcomposite-more-fine-grained}.

Given $k+b \geq 1$, we derive from (\ref{eqpd2}) and (\ref{eqn_expr_lahkk1case1}) that
\begin{align}
\hspace{-5pt}\mathbb{P}[\delta \leq k+b -2] \begin{cases}
\leq \sum_{h=0}^{k+b-2} \mathbb{P}[
\Phi_{n,h} \neq 0] \to 0,\textrm{ if }k+b \geq 2,\\
 =0,\textrm{ if }k+b =1,
\end{cases} \nonumber 
\end{align}
which implies
\begin{align}
\mathbb{P}[\delta \leq k+b -2] & = o(1). \label{eqn_delta_k2}
\end{align}

Given $k+b \geq 1$, we obtain from (\ref{eqpd1}) and (\ref{eqn_expr_lahkk1case3}) that
 \begin{align}
\mathbb{P}[\delta \geq k+b +1] &  \leq  \mathbb{P}[\Phi_{n,k+b} = 0] = o(1). \label{eqn_delta_k1}
\end{align}

Given $k+b \geq 1$, we show from (\ref{eqn_1mindel2}) and (\ref{eqn_expr_lahkk1case2}) that
 \begin{align}
\mathbb{P}[\delta \geq k+b  ] &  \leq  \mathbb{P}[\Phi_{n,k+b-1} = 0] \sim e^{-
\frac{e^{-\beta_n}}{(k+b-1)!}}, \label{eqn_delta_k1given1}
\end{align}
and
show from (\ref{eqn_1min}) and (\ref{eqn_expr_lahkk1case1}) that
 \begin{align}
&  \mathbb{P}[\delta \geq k+b  ] \nonumber \\ & \geq  \mathbb{P}[\Phi_{n,k+b-1} = 0] - \boldsymbol{1}[k+b\geq 2]\cdot \sum_{h=0}^{k+b-2}
\mathbb{P}[  \Phi_{n,h} \neq 0] \nonumber \\ & \sim e^{-
\frac{e^{-\beta_n}}{(k+b-1)!}}.\label{eqn_delta_k1given2}
\end{align}
 Then (\ref{eqn_delta_k1given1}) and (\ref{eqn_delta_k1given2}) together induce
%
\begin{align}
\mathbb{P}[\delta \geq k+b] & \sim e^{-
\frac{e^{-\beta_n}}{(k+b-1)!}} . \label{eqn_delta_k3}
\end{align}
From (\ref{eqn_delta_k2}) and (\ref{eqn_delta_k1}), we have
\begin{align}
 & \mathbb{P}[(\delta \neq k+b) \cap (\delta \neq k+b-1)]\nonumber \\ & =
\mathbb{P}[\delta \geq k+b +1] + \mathbb{P}[\delta \leq k+b -2]  = o(1); \label{eqn_delta_k4}
 \end{align}
from (\ref{eqn_delta_k1}) and (\ref{eqn_delta_k3}), we obtain
\begin{align}
\mathbb{P}[\delta = k+b] & = \mathbb{P}[\delta \geq k+b] -
\mathbb{P}[\delta \geq k+b +1] \nonumber \\ & \sim e^{-
\frac{e^{-\beta_n}}{(k+b-1)!}}  \nonumber \\ & \to
\begin{cases}
e^{- \frac{e^{-\beta ^*}}{(k+b-1)!}} , &\hspace{-8pt}\textrm{if $\lim_{n \to
\infty} \beta_n = \beta ^* \in (-\infty, \infty)$,} \\
1,  &\hspace{-8pt}\textrm{if $ \lim_{n \to \infty} \beta_n = \infty$}, \\
0, &\hspace{-8pt}\textrm{if $ \lim_{n \to \infty} \beta_n = - \infty$};
 \end{cases} \label{eqn_delta_k5}
 \end{align}
and from (\ref{eqn_delta_k4}) and (\ref{eqn_delta_k5}), we conclude
\begin{align}
 &  \mathbb{P}[\delta = k+b-1]  \nonumber \\ & = 1-
\mathbb{P}[(\delta \neq k+b) \cap (\delta \neq k+b-1)] -
\mathbb{P}[\delta = k+b]   \nonumber \\ &   \to
\begin{cases}
1 - e^{- \frac{e^{-\beta ^*}}{(k+b-1)!}} , &\textrm{if $\lim_{n \to
\infty} \beta_n = \beta ^* \in (-\infty, \infty)$,} \\
0,  &\textrm{if $ \lim_{n \to \infty} \beta_n = \infty$}, \\
1, &\textrm{if $ \lim_{n \to \infty} \beta_n = - \infty$}.
 \end{cases} \label{eqn_delta_k6}
 \end{align}

Properties \ding{173}--\ding{176} of Theorem \ref{thm:exact_qcomposite2} follow from
 (\ref{eqn_delta_k4})--(\ref{eqn_delta_k6}).

To summarize, We have used Theorem \ref{thm:exact_qcomposite2} (proved in Section \ref{sec_est} later) to establish Theorem \ref{thm:exact_qcomposite} under the additional condition $|\alpha_n| = o(\ln n)$, and to establish Theorem \ref{thm:exact_qcomposite-more-fine-grained}. In Appendix E of the full version \cite{full}, we explain that whenever Theorem \ref{thm:exact_qcomposite} with $|\alpha_n| = o(\ln n)$ holds, then Theorem \ref{thm:exact_qcomposite} regardless of $|\alpha_n| = o(\ln n)$. \qeda

\section{Establishing Theorem \ref{thm:exact_qcomposite2}} \label{sec_est}

\subsection{Method of Moments}

For $h = 0,1, \ldots$, with $\Phi_{n,h}$ counting the number of nodes
with degree $h$ in $\mathbb{G}_q$, we will show that
$\Phi_{n,h} $ asymptotically follows a Poisson distribution with mean
$\lambda_{n,h}$.  
 This is done by using the method of moments; specifically, in view of \cite[Theorem
2.13]{2008asymptotic}, we will obtain the desired result upon
establishing
\begin{align}
 \mathbb{P} [\textrm{Nodes }v_{1}, v_{2}, \ldots, v_{m}\textrm{ have
degree }h] &  \sim {\lambda_{n,h}}^m / n^m. \label{eqn_node_v12n}
\end{align}

Therefore, if Lemma \ref{LEM1} below holds, then the proof of
property (a) in Theorem \ref{thm:exact_qcomposite2} is completed; in
particular, we will have that for any integers $h \geq 0$ and $\ell
\geq 0$,
\begin{align}
 \mathbb{P}[\phi_h = \ell]
 & \sim (\ell !)^{-1}{\lambda_h} ^{\ell}e^{-\lambda_{n,h}} .
 \label{eqn_phihell}
 \end{align}

%
%

\begin{lem} \label{LEM1}
Given (\ref{peq1}) (i.e., $p_{e, q} = \frac{\ln  n \pm O(\ln \ln
n)}{n}$), $ K_n = \omega(1)$ and $\frac{{K_n}^2}{P_n} = o(1)$, then
for any integers $m \geq 1$ and $h \geq 0$, we have
\begin{align}
 &  \mathbb{P} [\textrm{Nodes }v_{1}, v_{2}, \ldots, v_{m}\textrm{ have
degree }h] \nonumber  \\
 & \quad\quad\quad\quad\quad\quad\quad\quad\quad\quad~
 \sim  (h!)^{-m}  (n p_{e, q})^{hm} e^{-m n p_{e, q}};\nonumber
\end{align}
i.e., (\ref{eqn_node_v12n}) follows with $\lambda_{n,h} $ set by
\begin{align}
 \lambda_{n,h} & = n (h!)^{-1}(n p_{e, q})^h e^{-n
p_{e, q}}. \label{eqn_labmdah}
 \end{align}

\end{lem}

Section \ref{secprf:lem_pos_exp} details the proof of Lemma
\ref{LEM1}. Given (\ref{peq1}), we obtain the following two results,
which are frequently used in the rest of the paper:
\begin{align}
p_{e, q} & \sim  \frac{\ln n}{n},\label{eq_pe_lnnn}
\end{align}
and
\begin{align}
p_{e, q} & \leq \frac{2\ln n}{n}\textrm{ for all $n$ sufficiently
large}. \label{eq_pe_upper}
\end{align}

\section{The Proof of Lemma \ref{LEM1}} \label{secprf:lem_pos_exp}

To start with, we consider several notation that will be used
throughout. We recall that ${C}_{i j}$ is the event that the
communication channel between distinct nodes $v_i$ and $v_j$ is {\em
on}. Then we set $\boldsymbol{1}[C_{ij}]$ as the indicator variable
of event ${C}_{i j}$ by
\begin{align}
 \hspace{-2pt} \boldsymbol{1}[C_{ij}]& \hspace{-2pt} :=  \hspace{-2pt} \begin{cases}
1,~ \textrm{if the
channel between }v_i\textrm{ and }v_j\textrm{ is \textit{on}}; \\
0,~ \textrm{if the channel between }v_i\textrm{ and }v_j\textrm{ is
\textit{off}}.
\end{cases} \nonumber
\end{align}
We denote by $\mathcal {C}_m$ a $\binom{m}{2}$-tuple consisting of
all possible $\boldsymbol{1}[C_{ij}]$ with $1 \leq i < j \leq m$ as
follows:
\begin{align}
\mathcal {C}_m : = ( &\boldsymbol{1}[C_{12}],
,\ldots,\boldsymbol{1}[C_{1m}],~~~\boldsymbol{1}[C_{23}],
,\ldots,\boldsymbol{1}[C_{2m}], \nonumber \\ &
  \boldsymbol{1}[C_{34}], \ldots,\boldsymbol{1}[C_{3m}],~~~\ldots,~~~
\boldsymbol{1}[C_{(m-1),m}]). \nonumber
\end{align}

Recalling $S_i$ as the key set on node $v_i$, we define a $m$-tuple
$\mathcal {T}_m$ through
\begin{align}
 \mathcal {T}_m &  : = (S_1, S_2, \ldots, S_m). \nonumber
\end{align}
Then we define $\mathcal {L}_m$ as
\begin{align}
\mathcal {L}_m & : = (\mathcal {C}_m, \mathcal {T}_m). \nonumber
\end{align}
With $\mathcal {L}_m$, we have the \emph{on}/\emph{off} states of
all channels between nodes $v_1, v_2, \ldots, v_m$ and the key sets
$S_1, S_2, \ldots, S_m$ on these $m$ nodes, so all edges between
these nodes in graph $\mathbb{G}_q$ are determined.

Let $\mathbb{C}_m, \mathbb{T}_m$ and $\mathbb{L}_m$ be the sets of
all possible $\mathcal {C}_m, \mathcal {T}_m$ and $\mathcal {L}_m$,
respectively. We define $\mathbb {L}_m^{(0)}$ such that
$\big(\mathcal {L}_m \in \mathbb {L}_m^{(0)}\big)$ is the event that
there is no edge between any two of nodes $ v_1, v_2, \ldots, v_m $;
i.e.,
\begin{align}
\mathbb{L}_m^{(0)} := \{\mathcal {L}_m  \boldsymbol{\mid} & ~(|S_i
\cap S_j| < q) \textrm{ \textit{or} }(\boldsymbol{1}[C_{ij}]  = 0),
 \nonumber \\ & \quad\quad~~~~~ \forall i, j\textrm{ with }1 \leq i < j \leq m.\}.
 \label{def_Lm0}
\end{align}

We define $N_i$ as the neighborhood set of node $v_i$ for
$i=1,2,\ldots,m$, and define the node set $M_{j_1 j_2 \ldots j_m}$
 for all $j_1, j_2, \ldots, j_m \in \{0,1\}$ by
\begin{align}
&  M_{j_1 j_2 \ldots j_m} \nonumber \\ &  :=
\mathlarger{\Bigg\{}w\mathlarger{\Bigg|}
\begin{array}{l}
 w \in \mathcal {V} \setminus \{v_1, v_2, \ldots, v_m\};\textrm{ and} \\
 \textrm{for }i=1,2,\ldots,m, \begin{cases} w \in
N_i\textrm{ if }j_i = 1; \\  w \notin N_i\textrm{ if }j_i
=0.\end{cases}
\end{array}\hspace{-11pt}\mathlarger{\Bigg\}}. \nonumber
\end{align}
Clearly, the sets $M_{j_1 j_2 \ldots j_m}$ for $j_1, j_2, \ldots,
j_m \in \{0,1\}$ are mutually disjoint. Setting $\mathcal {V}_m : =
\{v_1, v_2, \ldots, v_m\}$ and $\overline{\mathcal {V}_m} : =
\mathcal {V} \setminus \mathcal {V}_m $, we obtain
\begin{align}
\hspace{-5pt} \bigcup_{j_1, j_2, \ldots, j_m \in
\{0,1\}}\hspace{-5pt}|M_{j_1 j_2 \ldots j_m} |
 = \overline{\mathcal {V}_m}
,\label{eqn_nodevi_h3}
\end{align}
and
\begin{align}
\hspace{-5pt} \bigcup_{\begin{subarray}{c}j_1, j_2, \ldots, j_m \in \{0,1\}: \\
\sum_{i=1}^{m}j_i \geq 1.
\end{subarray}}\hspace{-5pt} |M_{j_1 j_2 \ldots j_m} | =
\bigg( \bigcup_{i=1}^m N_i \bigg) \hspace{1pt} \mathlarger{\cap}
\hspace{2pt} \overline{\mathcal {V}_m} .\label{eqn_nodevi_h2}
\end{align}
%
%
%



We define $2^m$-tuple $\mathcal {M}_m$ through\footnote{For a
non-negative integer $x$, the term $0^{x}$ is short for
$\underbrace{00 \ldots 0}_{\textrm{``}x\textrm{''} \textrm{ number
of ``}0\textrm{''}}$.}
\begin{align}
 \hspace{-1pt} \mathcal {M}_m &\hspace{-2pt} := \hspace{-2pt} \big( |
M_{j_1 j_2 \ldots j_m} |
\boldsymbol{\mid} j_1, j_2, \ldots, j_m \in \{0,1\} \big)  \nonumber  \\
& \hspace{-2pt} = \hspace{-2pt} \big( | M_{0^m} |, |M_{0^{m-1}1}|,
|M_{0^{m-2}1,0}| , |M_{0^{m-2}1,1}| , \ldots\big).\nonumber
\end{align}

%

Let $\mathcal {E}$ be the event that each of $v_1, v_2, \ldots, v_m$
has a degree of $h$. Given $\mathcal {L}_m \in \mathbb {L}_m $, we
define $\mathbb{M}_m(\mathcal {L}_m)$ as the set of $\mathcal {M}_m$
under the condition that $\mathcal {E}$ occurs. Then it's
straightforward to compute $\mathbb{P} [\mathcal {E}] $ via
\begin{align}
 \mathbb{P} [\mathcal {E}]  & = \hspace{-3pt}
\sum_{\begin{subarray}{c}\mathcal {L}_m^{*} \in \mathbb{L}_m, \\
\mathcal {M}_m^{*} \in \mathbb{M}_m (\mathcal {L}_m^{*}).
\end{subarray}} \hspace{-3pt} \mathbb{P}
\big[ \big( \mathcal {L}_m = \mathcal {L}_m^{*} \big) \cap \big(
\mathcal {M}_m = \mathcal {M}_m^{*} \big) \big].  \label{eqr_probe}
\end{align}

%
%


Given that event $\mathcal {E}$ happens, if any two of nodes $v_1,
v_2, \ldots, v_m$ do not have any common neighbor in
$\overline{\mathcal {V}_m}= \mathcal {V} \setminus \{v_1, v_2,
\ldots, v_m\}$,
then $\mathcal {M}_m$ is determined and denoted by $\mathcal
{M}_m^{(0)}$ which satisfies
%
%
\begin{align}
\begin{cases}
|M_{0^{i-1}, 1, 0^{m-i}}|   = h,& \textrm{for }i=1,2,\ldots,m; \\
|M_{j_1 j_2 \ldots j_m}|  = 0,&\textrm{for } \sum_{i=1}^m j_i >
1;\\|M_{0^m}|
  = n - m - hm .
\end{cases} \nonumber
\end{align}

By (\ref{eqr_probe}), we further write $\mathbb{P}
 [\mathcal {E} ]$ as the sum of
\begin{align}
 \sum_{\begin{subarray}{c}\mathcal {L}_m^{*} \in \mathbb{L}_m, \\
\mathcal {M}_m^{*} \in \mathbb{M}_m (\mathcal {L}_m^{*}): \\
   \left(\mathcal {L}_m^{*}
   \notin \mathbb{L}_m^{(0)}\right) \\
\textrm{ or }\left(\mathcal {M}_m^{*} \neq
\mathcal{M}_m^{(0)}\right)
\end{subarray}} \hspace{-2pt}
 \mathbb{P} \big[ \big( \mathcal {L}_m \hspace{-2pt} =
\hspace{-1pt} \mathcal {L}_m^{*} \big) \cap \big( \mathcal {M}_m
\hspace{-2pt} = \hspace{-1pt} \mathcal {M}_m^{*} \big)
\big]\label{term1}
\end{align}
and
\begin{align}
 & \mathbb{P} \big[ \big( \mathcal {L}_m \in \mathbb{L}_m^{(0)}
\big) \cap \big( \mathcal {M}_m = \mathcal{M}_m^{(0)} \big) \big].
\label{term2}
\end{align}
%
%
Consequently, Lemma \ref{LEM1} holds after we prove the following
Propositions 1 and 2. In the rest of the paper, we will often use
$1+x \leq e^x$ for any $x \in \mathbb{R}$ and $1 - xy \leq (1-x)^y
\leq 1 - xy + \frac{1}{2} x^2 y ^2$ for $0\leq x <1$ and $y = 0 , 1,
2, \ldots$ (Fact 2 in \cite{ZhaoYaganGligor}).

\begin{proposition}\label{PROP_ONE}
Given (\ref{peq1}) (i.e., $p_{e, q} = \frac{\ln  n \pm O(\ln \ln
n)}{n}$), $ K_n = \omega(1)$ and $\frac{{K_n}^2}{P_n} = o(1)$, we
have
\begin{align}
&  (\ref{term1}) =  o \left((h!)^{-m} (n p_{e, q})^{hm} e^{-m n
p_{e, q}}\right). \nonumber
\end{align}
\end{proposition}

\begin{proposition} \label{PROP_SND}
Given (\ref{peq1}) (i.e., $p_{e, q} = \frac{\ln  n \pm O(\ln \ln
n)}{n}$), $ K_n = \omega(1)$ and $\frac{{K_n}^2}{P_n} = o(1)$, we
have
\begin{align}
 &(\ref{term2}) \sim (h!)^{-m} (n p_{e, q})^{hm} e^{-m n p_{e, q}}. \nonumber
\end{align}
\end{proposition}

\section{The Proof of Proposition \ref{PROP_ONE}} \label{sec:PROP_ONE}

We embark on the evaluation of (\ref{term1}) by
computing
\begin{align}
\mathbb{P} \big[ \big( \mathcal {M}_m = \mathcal {M}_m^{*} \big)
\boldsymbol{\mid} \mathcal {L}_m = \mathcal {L}_m^{*} \big].
\label{eq_MmMm}
\end{align}
With $\mathcal {C}_m ^{*}$ and $\mathcal {T}_m ^{*}$ defined such
that $\mathcal {L}_m^{*} = (\mathcal {C}_m^{*}, \mathcal
{T}_m^{*})$, event $\big(\mathcal {L}_m = \mathcal {L}_m^{*}\big)$
is the union of events $\big(\mathcal {C}_m = \mathcal
{C}_m^{*}\big)$ and $\big(\mathcal {T}_m = \mathcal {T}_m^{*}\big)$.
Since $( \mathcal {C}_m \hspace{-1pt} = \hspace{-1pt} \mathcal
{C}_m^{*} )$ and $( \mathcal {M}_m \hspace{-1pt} = \hspace{-1pt}
\mathcal {M}_m^{*} )$ are independent, we obtain
\begin{align}
(\ref{eq_MmMm}) & = \mathbb{P} \big[ \big( \mathcal {M}_m = \mathcal
{M}_m^{*} \big)
 \boldsymbol{\mid}
\big( \mathcal {T}_m = \mathcal {T}_m^{*} \big) \big] \nonumber
.
\end{align}

For any $j_1, j_2, \ldots, j_m \in \{0,1\},$ for any distinct nodes
$w_1 \hspace{-2pt} \in \hspace{-2pt} \overline{\mathcal {V}_m} $ and
$ w_2 \hspace{-2pt} \in \hspace{-2pt} \overline{\mathcal {V}_m} $,
events $(w_1 \hspace{-2pt} \in \hspace{-2pt} M_{j_1 j_2 \ldots
j_m})$ and $(w_2 \in M_{j_1 j_2 \ldots j_m})$ are not independent
\cite{ryb3}, but are conditionally independent given $(\mathcal
{T}_m = \mathcal {T}_m^{*})$ (with the key sets $S_1, S_2, \ldots,
S_m$ specified as $S_1^{*}, S_2^{*}, \ldots, S_m^{*}$,
respectively). Therefore,
\begin{align}
 & \hspace{-2pt} (\ref{eq_MmMm}) = f(n-m , \mathcal {M}_m^{*})\mathbb{P} [w \in M_{0^m}^{*}
\hspace{-2pt} \boldsymbol{\mid} \hspace{-2pt} \mathcal
{T}_m = \mathcal {T}_m^{*} ]^{|M_{0^m}^{*} |} \times \nonumber  \\
& \hspace{-2pt} \prod_{\begin{subarray}{c}j_1, j_2, \ldots, j_m \in \{0,1\}: \\
\sum_{i=1}^{m}j_i \geq 1.
\end{subarray}} \hspace{-2pt} \mathbb{P}[w \in M_{j_1 j_2 \ldots j_m}^{*}
 \hspace{-2pt} \boldsymbol{\mid} \hspace{-2pt} \mathcal
{T}_m = \mathcal {T}_m^{*}]^{|M_{j_1 j_2 \ldots j_m}^{*} |},
\label{eqn_epsilonmmmm}
\end{align}
where $f\big(\sum_{i=1}^{\ell} x_i , (x_1, x_2, \ldots,
x_{\ell})\big)$ for integers $\ell \geq 1$ and $x_i \geq 0$ with $i
= 1,2 , \ldots, \ell$ is determined by
\begin{align}
&  f\bigg(\sum_{i=1}^{\ell} x_i , (x_1, x_2, \ldots,
x_{\ell})\bigg) \nonumber  \\
& \quad : = \binom{\sum_{i=1}^{\ell} x_i }{x_1 }
\binom{\sum_{i=2}^{\ell} x_i }{x_2 } \ldots
\binom{\sum_{i=\ell-1}^{\ell} x_i }{x_{\ell-1} }
\binom{x_{\ell} }{x_{\ell} } \nonumber  \\
& \quad \hspace{2pt} = \frac{\big(\sum_{i=1}^{\ell} x_i \big)!}{x_1
! x_2 ! \ldots x_{\ell} !}. \label{funcf}
\end{align}
From (\ref{funcf}) and
\begin{align}
\sum_{j_1, j_2, \ldots, j_m \in \{0,1\}}|M_{j_1 j_2 \ldots j_m}^{*}|
= n - m \label{eqn_nodevi_h5}
\end{align}
which holds by (\ref{eqn_nodevi_h3}), we have
\begin{align}
&   f(n-m  , \mathcal {M}_m^{*}) \nonumber  \\
& \hspace{-2pt} = \hspace{-2pt} \frac{ ( \sum_{j_1, j_2, \ldots, j_m
\in \{0,1\}}|M_{j_1 j_2 \ldots j_m}^{*}| ) !}{\prod_{j_1, j_2,
\ldots, j_m \in \{0,1\}}
(|M_{j_1 j_2 \ldots j_m}^{*}|! )} \nonumber  \\
&  \hspace{-2pt} = \hspace{-2pt} \frac{ (n \hspace{-2pt}
-\hspace{-2pt} m ) ! \hspace{-2pt} \Big/ \hspace{-2pt}
\Big(n\hspace{-2pt} -\hspace{-2pt} m - \hspace{-3pt}
\sum_{\begin{subarray}{c}j_1, j_2, \ldots, j_m \in \{0,1\}: \\
\sum_{i=1}^{m}j_i \geq 1.
\end{subarray}}|M_{j_1 j_2 \ldots j_m}^{*}|\hspace{-1pt}\Big)!}
{\prod_{\begin{subarray}{c}j_1, j_2, \ldots, j_m \in \{0,1\}: \\
\sum_{i=1}^{m}j_i \geq 1.
\end{subarray}} (|M_{j_1 j_2 \ldots j_m}^{*}|! )} \label{eqn_fnexpr}  \\
&  \hspace{-2pt} \leq \hspace{-2pt} n^{\sum_{\begin{subarray}{c}j_1,
j_2, \ldots, j_m
\in \{0,1\}: \\
\sum_{i=1}^{m}j_i \geq 1.
\end{subarray}}|M_{j_1 j_2 \ldots j_m}^{*}|}. \label{eqn_fnm}
\end{align}
Denoting $\sum_{\begin{subarray}{c}j_1, j_2, \ldots, j_m \in \{0,1\}: \\
\sum_{i=1}^{m}j_i \geq 1.
\end{subarray}}|M_{j_1 j_2 \ldots j_m}^{*}|$ by $\Lambda$, we prove
$\Lambda \leq hm - 1$ below if $\big(\mathcal {L}_m^{*} \notin
\mathbb{L}_m^{(0)} \big)$ or $\big(\mathcal {M}_m^{*} \neq
\mathcal{M}_m^{(0)} \big)$.

 On the one hand, assuming $\mathcal
{L}_m^{*} \notin \mathbb{L}_m^{(0)}$, there exist $i_1$ and $i_2$
with $1 \leq i_1 < i_2 \leq m$ such that nodes $v_{i_1}$ and
$v_{i_2}$ are neighbors with each other. Hence, $ \{v_{i_1},
v_{i_2}\} \subseteq [( \bigcup_{i=1}^m N_i ) \bigcap \mathcal {V}_m
]$. Then from (\ref{eqn_nodevi_h2}),
\begin{align}
 & \Lambda =
 \bigg|\bigcup_{i=1}^m N_i\bigg|  -
  \bigg|\bigg( \bigcup_{i=1}^{m} N_i \bigg) \hspace{2pt} \mathlarger{\cap} \hspace{2pt} \mathcal {V}_m\bigg|
 \leq hm - 2. \nonumber 
\end{align}

On the other hand, assuming $\mathcal {M}_m^{*} \neq
\mathcal{M}_m^{(0)}$, there exist $i_3 $ and $ i_4$ with $1 \leq i_3
< i_4 \leq m$ such that $N_{i_3} \cap N_{i_4} \neq \emptyset$. Then
from (\ref{eqn_nodevi_h2}),
\begin{align}
 & \Lambda \leq
 \bigg|\bigcup_{i=1}^m N_i\bigg|  \leq
  \bigg(\sum_{i=1}^m |N_i|\bigg) - |N_{i_3} \cap N_{i_4}| \leq
 hm - 1. \nonumber 
\end{align}

To summarize, if $\big(\mathcal {L}_m^{*} \notin \mathbb{L}_m^{(0)}
\big)$ or $\big(\mathcal {M}_m^{*} \neq \mathcal{M}_m^{(0)} \big)$,
we have
\begin{align}
\Lambda \leq hm - 1 ,   \label{lambda}
\end{align}
along with (\ref{eqn_nodevi_h5}) leading to
\begin{align}
| M_{0^m}^{*} |  &=  n - m -
 \Lambda > n - m - hm.  \label{eqn_M0m2n}
\end{align}

For any $j_1, j_2, \ldots, j_m \in \{0,1\}$ with $\sum_{i=1}^{m}j_i
\geq 1$, there exists $t \in \{0,1,\ldots, m\}$ such that $j_t = 1$,
so
\begin{align}
&  \mathbb{P}\big[w \in M_{j_1 j_2 \ldots j_m} \boldsymbol{\mid}
\mathcal {T}_m = \mathcal {T}_m^{*} \big]   \nonumber  \\
&  \leq  \mathbb{P}[E_{w v_t}  \boldsymbol{\mid} \mathcal {T}_m =
\mathcal {T}_m^{*}] = \mathbb{P}[E_{w v_t} ] = p_{e, q},
 \label{eqn_pe_not00}
\end{align}
where $E_{w v_t}$ is the event that there exists an edge between
nodes $w$ and $v_t$ in graph $\mathbb{G}_q$.

Substituting (\ref{eqn_fnm}-\ref{eqn_pe_not00}) into
(\ref{eqn_epsilonmmmm}), we obtain that if \\$\big(\mathcal
{L}_m^{*} \notin \mathbb{L}_m^{(0)} \big)$ or $\big(\mathcal
{M}_m^{*} \neq \mathcal{M}_m^{(0)} \big)$, then
\begin{align}
\hspace{-1pt}  (\ref{eq_MmMm})
 & <
 (np_{e,q})^{hm - 1}
\hspace{-1pt} \times \hspace{-1pt} \mathbb{P} [w \in M_{0^m}
\hspace{-1pt} \boldsymbol{\mid}  \hspace{-1pt} \mathcal {T}_m =
\mathcal {T}_m^{*} ]^{n - m - hm} .  \label{eq_pmmll2}
\end{align}

Applying (\ref{eq_MmMm}) and (\ref{eq_pmmll2}) to (\ref{term1}),
 we
get
\begin{align}
(\ref{term1}) & < \sum_{\mathcal {L}_m^{*} \in \mathbb{L}_m}
\Big\{ |\mathbb{M}_m (\mathcal {L}_m^{*})|   \nonumber  \\
& \quad \times \textrm{R.H.S. of (\ref{eq_pmmll2})} \times
\mathbb{P} \big[ \mathcal {L}_m = \mathcal {L}_m^{*} \big] \Big\}.
  \label{eqn_TmCmt}
\end{align}


%
%
To bound $|\mathbb{M}_m (\mathcal {L}_m^{*})|$, note that $\mathcal
{M}_m$ is a $2^m$-tuple. Among the $ 2^m $ elements of the tuple,
each of $|M_{j_1 j_2 \ldots j_m}
|\big|_{\begin{subarray}{c}j_1, j_2, \ldots, j_m \in \{0,1\}: \\
\sum_{i=1}^{m}j_i \geq 1.
\end{subarray}}$ is at least 0 and at most $h$; and the remaining
element $| M_{0^m} |$ can be determined by (\ref{eqn_nodevi_h5}).
Then it's straightforward that
\begin{align}
|\mathbb{M}_m (\mathcal {L}_m^{*})| &  \leq (h+1)^{2^m-1}.
\label{eqn_MmLm}
\end{align}

Using (\ref{eqn_MmLm}) in (\ref{eqn_TmCmt}), and considering
$\big(\mathcal {L}_m = \mathcal {L}_m^{*}\big)$ is the union of
independent events $\big(\mathcal {T}_m = \mathcal {T}_m^{*}\big)$
and $\big(\mathcal {C}_m \hspace{-1pt} = \hspace{-1pt} \mathcal
{C}_m^{*}\big)$, and $\sum_{\mathcal {C}_m^{*} \in \mathbb{C}_m}
\hspace{-1pt} \mathbb{P} \big[ \mathcal {C}_m \hspace{-1pt} =
\hspace{-1pt} \mathcal {C}_m^{*} \big] \hspace{-1pt} = \hspace{-1pt}
1$, we derive
\begin{align}
(\ref{term1}) &  <  (h+1)^{2^m-1} (np_{e,q})^{hm-1} \hspace{-2pt}
\times \hspace{-2pt}\sum_{\mathcal {T}_m^{*} \in \mathbb{T}_m}
\hspace{-4pt} \Big\{
\mathbb{P}\big[ \mathcal {T}_m = \mathcal {T}_m^{*} \big]   \nonumber  \\
& \quad \times \mathbb{P} [w \in M_{0^m} \boldsymbol{\mid} \mathcal
{T}_m = \mathcal {T}_m^{*} ]^{n - m - hm} \Big\} .
\label{prop_prf}
\end{align}
From (\ref{prop_prf}) and $\lim_{n \to \infty} n p_{e, q} = \infty $
by (\ref{eq_pe_lnnn}), the proof of Proposition \ref{PROP_ONE} is
completed once we show
\begin{align}
 & \sum_{ \mathcal {T}_m^{*} \in \mathbb{T}_m } \mathbb{P}[\mathcal {T}_m = \mathcal {T}_m^{*}]
\mathbb{P} [w \in M_{0^m} \boldsymbol{\mid} \mathcal
{T}_m = \mathcal {T}_m^{*} ]^{n - m - hm}   \nonumber  \\
& \quad  \leq e^{- m n p_{e, q}} \cdot [1+o(1)] .  \label{EQ}
\end{align}

\subsection{Establishing (\ref{EQ})}

From (\ref{eq_evalprob_3_qcmp}) and (\ref{eq_evalprob_4_qcmp})
(viz., Lemma \ref{lem_evalprob_qcmp} in the Appendix), it holds that
\begin{align}
 & \mathbb{P} [w \in M_{0^m}^{*} \boldsymbol{\mid} \mathcal
{T}_m = \mathcal {T}_m^{*} ]^{n - m - hm } \nonumber  \\
& =  \mathbb{P} [w  \in M_{0^m}^{*}   \boldsymbol{\mid}
 \mathcal {T}_m   = \mathcal {T}_m^{*} ]^{
n} \mathbb{P} [w \in   M_{0^m}^{*}   \boldsymbol{\mid}
 \mathcal {T}_m =   \mathcal {T}_m^{*} ]^{-m - h m}  \nonumber  \\
&   \leq   e^{- m n p_{e, q}   +   (q+2)! \binom{m}{2} n{(p_{e,
q})}^{\frac{q+1}{q}}  +
  \frac{n p_{e, q} p_n}{K_n}
 \sum_{1\leq i <j \leq m} |S_{ij}^{*}|}
 \nonumber  \\
& \quad \times (1 - m p_{e, q})^{-m - h m},  \label{eqn_prbwM}
\end{align}
where $S_{ij}^{*} = S_{i}^{*} \cap S_{j}^{*}$. With
(\ref{eq_pe_lnnn}) (i.e., $p_{e, q} \sim \frac{\ln n}{n}$), we have
$m^2 n {p_{e, q}}^2
  = o(1)$ and $m p_{e,q} =
o(1)$, which are substituted into (\ref{eqn_prbwM}) to induce
(\ref{EQ}) once we prove%
%
%
\begin{align}
\sum_{ \mathcal {T}_m^{*} \in \mathbb{T}_m  } \mathbb{P}[\mathcal
{T}_m = \mathcal {T}_m^{*}] e^{\frac{n p_{e, q} p_n}{K_n}\sum_{1\leq
i <j \leq m}|S_{ij}^{*}|} &  \leq 1+o(1). \label{eqn_sumTmst}
\end{align}
 L.H.S. of (\ref{eqn_sumTmst}) is denoted by $H_{n,m}$ and evaluated
below. For each fixed and sufficiently large $n$, we consider: {a)}
{${ p_n <  n^{-\delta} (\ln n)^{-1}}$} and {b)} {${ p_n \geq
n^{-\delta} (\ln n)^{-1}}$}, where $\delta$ is an arbitrary constant
with $0<\delta<1$.

\noindent \textbf{a)} $\boldsymbol{ p_n <  n^{-\delta} (\ln
n)^{-1}}$

From $p_n < n^{-\delta} (\ln n)^{-1}$, (\ref{eq_pe_upper}) (namely,
$p_{e,q} \leq \frac{2\ln n}{n}$) and $|S_{ij}^{*}| \leq K_n$ for
$1\leq i <j \leq m$, it holds that
\begin{align}
e^{\frac{n p_{e,q} p_n}{K_n} \sum_{i =1}^{m-1}|S_{i  m}^*|} & < e^{2
\ln n \cdot n^{-\delta} (\ln n)^{-1} \cdot \binom{m}{2}} <  e^{ m^2
n^{-\delta}},\nonumber
\end{align}
which is substituted into $H_{n,m}$ to bring about
\begin{align}
& H_{n,m} < e^{ m^2 n^{-\delta}} \sum_{ \mathcal {T}_m^{*} \in
\mathbb{T}_m  } \mathbb{P}[\mathcal {T}_m = \mathcal {T}_m^{*}] =
e^{ m^2 n^{-\delta}} , \nonumber
\end{align}

%
%

%

\noindent \textbf{b)} $\boldsymbol{ p_n \geq  n^{-\delta} (\ln
n)^{-1}}$

We relate $H_{n,m}$ to $H_{n,m-1}$ and assess $H_{n,m}$ iteratively.
First, with $\mathcal {T}_m^{*} = (S_1^*, S_2^*, \ldots, S_m^*)$,
event $(\mathcal {T}_m = \mathcal {T}_m^{*})$ is the intersection of
independent events: $(\mathcal {T}_{m-1} = \mathcal {T}_{m-1}^{*})$
and $(S_m = S_m^*)$. Then we have
\begin{align}
& H_{n,m}   \nonumber \\
&  = \sum_{ \begin{subarray} ~\mathcal {T}_{m-1}^* \in
\mathbb{T}_{m-1} , \\  \hspace{9pt}S_m^* \in \mathbb{S}_m
\end{subarray} } \Big( \mathbb{P}[(\mathcal {T}_{m-1} = \mathcal
{T}_{m-1}^{*}) \cap (S_m =
S_m^*)] \times  \nonumber \\
&   \quad\quad \quad\quad e^{\frac{n p_{e,q} p_n}{K_n} \sum_{1\leq i
<j \leq m-1}|S_{ij}^*|} e^{\frac{n p_{e,q} {p_n} }{K_n} \sum_{i
=1}^{m-1}|S_{i  m}^*|} \Big)  \nonumber \\ &  = H_{n,m-1} \cdot
\sum_{S_m^* \in \mathbb{S}_m} \mathbb{P}[ S_m = S_m^* ]  e^{\frac{n
p_{e,q} p_n}{K_n} \sum_{i =1}^{m-1}|S_{i  m}^*|} . \label{HnmHnm1}
\end{align}
By $ \sum_{i =1}^{m-1} \hspace{-2pt} |S_{i  m}^*| \hspace{-3pt} =
\hspace{-3pt} \sum_{i =1}^{m-1} \hspace{-2pt} |S_i^* \cap S_m^*|
\hspace{-3pt} \leq \hspace{-3pt} m \big|S_m^* \hspace{-1pt} \cap
\hspace{-1pt}
 \big(\hspace{-2pt}\bigcup_{i =1}^{m-1} \hspace{-2pt}S_{i }^* \big) \hspace{-1pt} \big|$ and (\ref{eq_pe_upper})
 (i.e., $ p_{e,q} \leq  \frac{2\ln n}{n}$), we get
\begin{align}
&  e^{\frac{n p_{e,q} p_n }{K_n} \sum_{i =1}^{m-1}|S_{i  m}^*|} \leq
e^{ \frac{2m p_n \ln n}{K_n} |S_m^* \cap
 (\bigcup_{i =1}^{m-1}S_{i }^* ) |} ,
\nonumber
\end{align}
further leading to
\begin{align}
&  H_{n,m} / H_{n,m-1} \nonumber
\\ 
&  \leq \sum_{u=0}^{K_n} \mathbb{P}\bigg[\bigg|S_m^* \bigcap
\bigg(\bigcup_{i =1}^{m-1}S_{i }^*\bigg)\bigg| = u \bigg] e^{\frac{2
u m  {p_n} \ln n}{K_n}} .\label{eqn_tmtm-1}
\end{align}
%
Denoting $\big|\bigcup_{i=1}^{m-1}S_{i}^*\big|$ by $v$, then we
obtain that for $u \in [\max\{0, K_n + v - P_n\}  , K_n] $,
\begin{align}
 \mathbb{P}\bigg[\bigg|S_m^* \bigcap
\bigg(\bigcup_{i=1}^{m-1}S_{i}^*\bigg)\bigg| = u \bigg] &  =
\frac{\binom{v}{u} \binom{P_n - v}{K_n - u}}{\binom{P_n}{K_n}},
\label{probsm}
\end{align}
which together with $ K_n \leq  v \leq m K_n$ yields
\begin{align}
& \textrm{L.H.S. of (\ref{probsm})} \nonumber
\\& \quad \leq \frac{(m K_n)^u}{u!} \cdot
 \frac{(P_n - K_n)^{K_n - u}}{(K_n - u)!} \cdot \frac{K_n !}{(P_n - K_n)^{K_n}}
\nonumber
\\& \quad \leq \frac{1}{u!} \bigg( \frac{m {K_n}^2}{P_n - K_n}\bigg)^u. \label{probsm2}
\end{align}
For $u \notin [\max\{0, K_n + v - P_n\}  , K_n] $, L.H.S. of
(\ref{probsm}) equals 0. Then from (\ref{eqn_tmtm-1}) and
(\ref{probsm2}),
\begin{align}
\textrm{R.H.S. of (\ref{eqn_tmtm-1})} &  \leq  \sum_{u=0}^{K_n}
\frac{1}{u!} \bigg( \frac{m {K_n}^2}{P_n - K_n} \cdot e^{\frac{2 m
{p_n} \ln n}{K_n}}\bigg)^u \nonumber
\\& \quad \leq  e^{\frac{m {K_n}^2}{P_n - K_n} \cdot e^{\frac{2 m  {p_n} \ln
n}{K_n}}}. \label{umKnPN}
\end{align}

By $\frac{{K_n}^2}{P_n} = o(1)$ and Lemma \ref{lem_eval_psq}-Property (i),
\begin{align}
\frac{{K_n}^2}{P_n-K_n}
  & \leq \frac{{K_n}^2}{P_n}  \cdot [1+o(1)]
 \leq \big( q! p_{s,q} \big)^{\frac{1}{q}} \cdot [1+o(1)] .\label{eqn_knpn_qcmp}
\end{align}
For $n$ sufficiently large, from $p_n \geq n^{-\delta} (\ln n)^{-1}$
and (\ref{eq_pe_upper})
  (i.e., $p_{e,q} =p_n p_{s,q}   \leq \frac{2\ln n}{n}$), we have
\begin{align}
p_{s,q}   & =  {p_n} ^{-1} {p_{e,q}} \leq {p_n} ^{-1} \cdot
2n^{-1}\ln n \leq 2 n^{\delta-1} (\ln n)^2. \label{eqps0}
\end{align}

From (\ref{eqn_knpn_qcmp}) and (\ref{eqps0}),
\begin{align}
\frac{{K_n}^2}{P_n-K_n}
  & \leq [ q! \cdot 2 n^{\delta-1} (\ln n)^2]^{\frac{1}{q}}
  \cdot [1+o(1)]  \nonumber
\\& \leq 3 q \cdot
n^{\frac{\delta-1}{q}} (\ln n)^{\frac{2}{q}}. \label{eqn_knpn2x}
\end{align}

Given $K_n =  \omega(1)  $, for arbitrary constant $c > q$ and for
all $n$ sufficiently large, $\frac{K_n}{p_n} \geq \frac{2cq \cdot
m}{(c-q)(1-\delta)} $ holds. Then
\begin{align}
e^{\frac{2 m p_n \ln n}{K_n}} & \leq e^{  \frac{(c-q)(1-\delta)}{cq}
\ln n} = n^{\frac{(c-q)(1-\delta)}{cq}}  .\label{ja1}
\end{align}
The use of (\ref{umKnPN}) (\ref{eqn_knpn2x}) and (\ref{ja1}) in
(\ref{eqn_tmtm-1}) yields
\begin{align}
 & H_{n,m} / H_{n,m-1} \leq \textrm{R.H.S. of (\ref{eqn_tmtm-1})}
 \nonumber \\ & \leq  e^{ 3 qm \cdot
n^{\frac{\delta-1}{q}} (\ln n)^{\frac{2}{q}} \cdot
n^{\frac{(c-q)(1-\delta)}{cq}} } \leq \Big(e^{3 q \cdot
n^{\frac{\delta-1}{c}} (\ln n)^{\frac{2}{q}}} \Big)^m.
\label{gnmgnm-1}
\end{align}

To derive $H_{n,m}$ iteratively based on (\ref{gnmgnm-1}), we
compute $H_{n,2}$ below. By definition, setting $m=2$ in L.H.S. of
(\ref{eqn_sumTmst}) and considering the independence between events
$(S_1  = S_1^*)$ and $(S_2  = S_2^*)$, we gain
\begin{align}
  H_{n,2} &   =    \sum_{S_1^*
\in \mathbb{S}_m}   \mathbb{P}[ S_1  = S_1^* ]
   \sum_{S_2^* \in \mathbb{S}_m}  \mathbb{P}[ S_2  =
S_2^* ] e^{\frac{n p_{e,q} p_n}{K_n} |S_1^* \cap S_2^*|}.
\label{eqn_gn2}
\end{align}
Clearly, $\sum_{S_2^* \in \mathbb{S}_m} \hspace{-3pt} \mathbb{P}[
S_2 \hspace{-1pt} = \hspace{-1pt} S_2^* ] e^{\frac{n p_{e,q}
p_n}{K_n} |S_1^* \cap S_2^*|} $ equals R.H.S. of (\ref{eqn_tmtm-1})
with $m = 2$. Then from (\ref{gnmgnm-1}) and (\ref{eqn_gn2}),
\begin{align}
H_{n,2}  &  \leq \sum_{S_1^* \in \mathbb{S}_m} \mathbb{P}[ S_1  =
S_1^* ]  e^{6 q \cdot n^{\frac{\delta-1}{c}} (\ln n)^{\frac{2}{q}}}
=  e^{6 q \cdot n^{\frac{\delta-1}{c}} (\ln n)^{\frac{2}{q}}}.
\label{hn2}
\end{align}

Therefore, it holds via (\ref{gnmgnm-1}) and (\ref{hn2}) that
\begin{align}
H_{n,m}  & \leq \Big(e^{3 q \cdot n^{\frac{\delta-1}{c}} (\ln
n)^{\frac{2}{q}}}  \Big)^{m+(m-1) + \ldots + 3} \cdot e^{6 q \cdot n^{\frac{\delta-1}{c}}
 (\ln n)^{\frac{2}{q}}} \nonumber \\
&  = e^{\frac{3}{2}q(m^2+m-2) n^{\frac{\delta-1}{c}} (\ln
n)^{\frac{2}{q}}} \nonumber.
\end{align}

Finally, summarizing cases a) and b), we report
\begin{align}
 H_{n,m} & \leq  \max\left\{e^{ m^2 n^{-\delta}} ,
e^{\frac{3}{2}q(m^2+m-2) n^{\frac{\delta-1}{c}} (\ln
n)^{\frac{2}{q}}}\right\} . \nonumber
\end{align}
With $n \to \infty$, $ H_{n,m} \leq 1+ o(1)$ (i.e.,
(\ref{eqn_sumTmst})) follows.

\section{The Proof of Proposition \ref{PROP_SND}} \label{sec:PROP_SND}

%
%
%
%

We define $\mathcal{C}_m^{(0)}$ and $\mathbb{T}_m^{(0)}$ by
\begin{align}
\mathcal {C}_m^{(0)} & = ( \underbrace{0, 0, \ldots,
0}_{\binom{m}{2} \textrm{ number of ``}0\textrm{''}} ), \nonumber
\end{align}
and
\begin{align}
\mathbb{T}_m^{(0)}
 & =
 \{\mathcal {T}_m
 \boldsymbol{\mid}
| S_i \cap S_j | < q, ~ \forall i, j\textrm{ with }1 \leq i < j \leq
m.\}. \nonumber
\end{align}
Clearly, $\big(\mathcal {C}_m = \mathcal{C}_m^{(0)}\big)$ or
$\big(\mathcal {T}_m \in \mathbb{T}_m^{(0)}\big)$ each implies
$\big( \mathcal {L}_m \in \mathbb{L}_m^{(0)} \big)$. Also,
$\big(\mathcal {C}_m = \mathcal{C}_m^{(0)}\big)$ and $\big(\mathcal
{M}_m = \mathcal{M}_m^{(0)}\big)$ are independent with each other.
Therefore, with $(\ref{term2}) = \mathbb{P} \big[ \big( \mathcal
{L}_m \in \mathbb{L}_m^{(0)} \big) \cap \big( \mathcal {M}_m =
\mathcal{M}_m^{(0)} \big) \big]$, we get
\begin{align}
& (\ref{term2}) \geq \mathbb{P} \big[ \mathcal {C}_m =
\mathcal{C}_m^{(0)}\big] \mathbb{P} \big[ \mathcal {M}_m =
\mathcal{M}_m^{(0)} \big], \label{prcm}
\end{align}
and
\begin{align}
& (\ref{term2}) \geq \mathbb{P} \big[ \mathcal {T}_m \hspace{-1pt}
\in  \hspace{-1pt} \mathbb{T}_m^{(0)}
 \big] \mathbb{P} \big[ \big( \mathcal {M}_m  \hspace{-1pt} =  \hspace{-1pt} \mathcal{M}_m^{(0)} \big) \hspace{-2pt}
  \boldsymbol{\mid} \hspace{-2pt}
 \big( \mathcal {T}_m  \hspace{-1pt} \in
\mathbb{T}_m^{(0)}  \hspace{-1pt} \big)\big]. \label{eqn_tmtmst}
\end{align}

Given $\big(\mathcal {C}_m = \mathcal{C}_m^{(0)}\big) =
\overline{\bigcup_{ 1 \leq i < j \leq m} {C_{ij}}} $ and
\\$\big(\mathcal {T}_m \in \mathbb{T}_m^{(0)}\big)  = \overline{\bigcup_{ 1
\leq i < j \leq m} {\Gamma_{ij}} }$, applying the union bound, we
obtain
\begin{align}
 & \mathbb{P} \big[ \mathcal {C}_m = \mathcal{C}_m^{(0)} \big]\geq 1 -
\sum_{ 1 \leq i < j \leq m}\mathbb{P}[ C_{ij} ] \geq 1- m^2 p_n / 2,
\label{prcmpn}
\end{align}
and
\begin{align}
& \mathbb{P}\big[\mathcal {T}_m  \in \mathbb{T}_m^{(0)}\big] \geq 1
- \sum_{ 1 \leq i < j \leq m}\mathbb{P}[ \Gamma_{ij} ] \geq  1 - m^2
p_{s,q} / 2.\label{mthbbP}
\end{align}

In the following two subsections, we will prove
\begin{align}
\mathbb{P}  \big[ \mathcal {M}_m = \mathcal{M}_m^{(0)} \big] & \sim
(h!)^{-m} (n p_{e,q})^{hm} e^{-m n p_{e,q}} \label{eqn_prMm},
\end{align}
and
\begin{align}
&  \mathbb{P} \big[ \big( \mathcal {M}_m = \mathcal{M}_m^{(0)} \big)
\boldsymbol{\mid} \big( \mathcal {T}_m \in \mathbb{T}_m^{(0)}
\big)\big] \nonumber  \\
& \quad \geq  (h!)^{-m} (n p_{e,q})^{hm} e^{-m n p_{e,q}} \cdot
[1-o(1)] .\label{prob_MmMm_sim}
\end{align}

Substituting (\ref{prcmpn}) and (\ref{eqn_prMm}) into (\ref{prcm}),
and applying (\ref{mthbbP}) and (\ref{prob_MmMm_sim}) to
(\ref{eqn_tmtmst}), we have
\begin{align}
& \frac{(\ref{term2})}{(h!)^{-m} (n p_{e,q})^{hm} e^{-m n p_{e,q}}}  \nonumber  \\
& ~  \geq (  1 - \min\{ p_{s,q}, p_n \}  \cdot m^2  / 2)\cdot
[1-o(1)]
  . \label{pro2_pt1}
\end{align}
From (\ref{eqn_prMm}), we get
\begin{align}
 (\ref{term2}) &  \leq \mathbb{P} \big[ \mathcal {M}_m
\in
\mathbb{M}_m^{(0)} \big] \nonumber  \\
& \leq  (h!)^{-m} (n p_{e,q})^{hm} e^{-m n p_{e,q}} \cdot [1+o(1)].
\label{pro2_pr}
\end{align}
Combining (\ref{pro2_pt1}) and (\ref{pro2_pr}), and using $\min\{
p_{s,q}, p_n \} \leq \sqrt{p_{s,q} p_n}  = \sqrt{p_{e,q}} \leq
\sqrt{\frac{2\ln n}{n}} = o(1)$ which holds from $p_{e,q} = p_{s,q}
p_n$ and (\ref{eq_pe_upper}), Proposition 2 follows. Below we detail
the proofs of (\ref{eqn_prMm}) and (\ref{prob_MmMm_sim}).

%

\subsection{Establishing (\ref{eqn_prMm})}

We have
\begin{align}
&\mathbb{P} \big[ \mathcal {M}_m = \mathcal{M}_m^{(0)} \big]
 \nonumber  \\
& \sum_{\mathcal {T}_m^{*} \in \mathbb{T}_m}  \Big\{ \mathbb{P}
\big[ \mathcal {T}_m   =  \mathcal {T}_m^{*} \big] \mathbb{P} \big[
\big( \mathcal {M}_m   =  \mathcal{M}_m^{(0)} \big)
\boldsymbol{\mid} \big( \mathcal {T}_m   =   \mathcal {T}_m^{*}
\big)\big] \Big\},\nonumber
\end{align}
where
\begin{align}
& \mathbb{P} \big[ \big( \mathcal {M}_m = \mathcal{M}_m^{(0)}
 \big) \boldsymbol{\mid} \big( \mathcal {T}_m
= \mathcal {T}_m^{*} \big)\big] \nonumber  \\ & =  f\big(n-m ,
\mathcal{M}_m^{(0)}\big) \mathbb{P} [w \in M_{0^m}
\boldsymbol{\mid}
\mathcal {T}_m = \mathcal {T}_m^{*} ]^{n-m-hm} \nonumber  \\
& \quad \quad \times \prod_{i=1}^{m} \mathbb{P}[w \in
M_{0^{i-1}, 1, 0^{m-i}} \boldsymbol{\mid}
\mathcal {T}_m = \mathcal {T}_m^{*} ]^{h}, \label{pMmexpr}
\end{align}
with function $f $ specified in (\ref{funcf}). From
(\ref{eqn_fnexpr}),
\begin{align}
&   f\big(n\hspace{-1pt}-\hspace{-1pt}m , \mathcal{M}_m^{(0)}
\hspace{-1pt}\big) \hspace{-2pt} = \hspace{-2pt} \frac{ (n
\hspace{-1pt}- \hspace{-1pt}m) ! }{(n \hspace{-1pt}-\hspace{-1pt}
m\hspace{-1pt} -\hspace{-1pt} hm)!(h!)^m} \hspace{-3pt} \sim
\hspace{-2pt} (h!)^{-m}n^{hm}. \label{eqn_f00}
\end{align}
 We will establish
\begin{align}
 &\hspace{-4pt}
  \sum_{ \mathcal {T}_m^{*} \in \mathbb{T}_m }
   \hspace{-7pt}\Big\{ \hspace{-1pt}
\mathbb{P}[\mathcal {T}_m \hspace{-2pt}
 = \hspace{-2pt}
 \mathcal {T}_m^{*}] \hspace{-2pt}
\prod_{i=1}^{m} \{ \mathbb{P}\big[w \hspace{-2pt}
 \in \hspace{-2pt}
 M_{0^{i-1}, 1,
0^{m-i}}^{(0)}  \hspace{-2pt}\boldsymbol{\mid} \hspace{-2pt}
\mathcal {T}_m \hspace{-2pt}
 = \hspace{-2pt}
 \mathcal {T}_m^{*} \big]^h  \} \hspace{-2pt}\Big\}
\nonumber  \\
& \quad  \geq {p_{e,q}}^{hm} \cdot [1-o(1)]
.\label{eq_evalprob_exp_2}
\end{align}
We use (\ref{eqn_f00}) and (\ref{eq_evalprob_exp_2}) as well as
(\ref{eq_evalprob_3_qcmp}) (viz., Lemma \ref{lem_evalprob_qcmp} in
the Appendix) in evaluating $\mathbb{P} \big[ \mathcal {M}_m =
\mathcal{M}_m^{(0)} \big]$ above. Then
\begin{align}
&  \mathbb{P} \big[ \mathcal {M}_m = \mathcal{M}_m^{(0)} \big]
\nonumber  \\
& \geq  (h!)^{-m}n^{hm} \cdot [1-o(1)] \cdot (1-m p_{e,q})^{n}
\times
 \nonumber  \\
&  \hspace{-3pt} \sum_{\mathcal {T}_m^{*} \in \mathbb{T}_m}
\hspace{-4pt} \mathbb{P}[\mathcal {T}_m \hspace{-2pt}
 = \hspace{-2pt}
 \mathcal {T}_m^{*}]
\prod_{i=1}^{m}\big\{ \mathbb{P}[w  \hspace{-2pt}  \in \hspace{-2pt}
 M_{0^{i-1},
1, 0^{m-i}} \hspace{-2pt}
 \boldsymbol{\mid} \hspace{-2pt}
 \mathcal {T}_m \hspace{-2pt}
 =  \hspace{-2pt}
 \mathcal {T}_m^{*} ]^{h} \big\} \nonumber \\
& \geq  (h!)^{-m} (n p_{e,q})^{hm} e^{-m n p_{e,q}} \cdot [1-o(1)] .
  \label{eqn_prMm_pt2}
\end{align}

Substituting (\ref{EQ}) (\ref{eqn_f00}) above
 and (\ref{eq_evalprob_1_qcmp}) in Lemma \ref{lem_evalprob_qcmp} into the computation of $\mathbb{P} \big[ \mathcal {M}_m = \mathcal{M}_m^{(0)}
\big]$ yields
\begin{align}
& \mathbb{P} \big[ \mathcal {M}_m = \mathcal{M}_m^{(0)} \big]
\nonumber  \\
& \leq  (h!)^{-m}n^{hm} {p_{e,q}}^{hm} \times [1+o(1)] \times \nonumber  \\
& \sum_{\mathcal {T}_m^{*} \in \mathbb{T}_m} \mathbb{P}[\mathcal
{T}_m = \mathcal {T}_m^{*}] \mathbb{P} [w \in M_{0^m}
\boldsymbol{\mid} \mathcal {T}_m = \mathcal {T}_m^{*}
]^{n-m-hm}  \nonumber \\
& \sim (h!)^{-m} (n p_{e,q})^{hm} e^{-m n p_{e,q}} .
\label{eqn_prMm_pt1}
\end{align}

Then (\ref{eqn_prMm}) follows from (\ref{eqn_prMm_pt2}) and
(\ref{eqn_prMm_pt1}). Namely, (\ref{eqn_prMm}) holds upon the
establishment of (\ref{eq_evalprob_exp_2}), which is proved below.
First, from (\ref{eq_evalprob_2_qcmp}) in Lemma
\ref{lem_evalprob_qcmp}, with $\mathcal {T}_m^{*} = (S_1^{*} ,
S_2^{*}  , \ldots, S_m^{*} ) $ and $S_{ij}^{*} = S_{i}^{*} \cap
S_{j}^{*}$, we get
\begin{align}
&  \hspace{-2pt} \prod_{i=1}^{m} \mathbb{P}\big[w \in M_{0^{i-1}, 1,
0^{m-i}}^{(0)} \boldsymbol{\mid} \mathcal {T}_m = \mathcal {T}_m^{*}
\big]^h
\nonumber  \\
&  \hspace{-2pt} \geq  \hspace{-3pt} { {p_{e,q}}^{hm} \hspace{-2pt}
\prod_{i=1}^{m}  \hspace{-2pt} \bigg[  \hspace{-2pt} 1 \hspace{-2pt}
- \hspace{-3pt} \bigg( \hspace{-2pt}
(q+2)!m{(p_{e,q})}^{\frac{1}{q}} \hspace{-2pt}
 + \hspace{-2pt} \frac{p_n}{K_n} \hspace{-3pt}
\sum_{j\in\{1,2,\ldots,m\}\setminus\{i\}}
\hspace{-4pt}  |S_{ij}^{*}|\hspace{-2pt} \bigg)\hspace{-2pt} \bigg]\hspace{-2pt}  }^h\nonumber  \\
& \hspace{-2pt} \geq \hspace{-3pt} {p_{e,q}}^{hm} \bigg( 1 - (q+2)!
h m^2 (p_{e,q})^{\frac{1}{q}}  - \frac{2 hp_n}{K_n} \sum _{1\leq i
<j \leq m} |S_{ij}^{*}|\bigg). \nonumber
\end{align}
With $p_{e,q} = o(1)$ by (\ref{eq_pe_lnnn}), we obtain
(\ref{eq_evalprob_exp_2}) once proving
\begin{align}
\frac{ p_n}{K_n} \hspace{-2pt} \sum_{ \mathcal {T}_m^{*} \in
\mathbb{T}_m } \hspace{-2pt} \hspace{-2pt} \bigg(
\mathbb{P}[\mathcal {T}_m = \mathcal {T}_m^{*}] \hspace{-2pt} \sum
_{1\leq i <j \leq m} \hspace{-2pt}|S_{ij}^{*}| \hspace{-1pt} \bigg)
& = o(1). \hspace{-2pt} \label{prfhpnkn}
\end{align}
Clearly, $| S_{ij}^{*} | \leq K_n$. If $\mathcal {T}_m^{*}   \in
\mathbb{T}_m^{(0)}$, it further holds that $| S_{ij}^{*} |  <   q $.
 Consequently, from (\ref{mthbbP}), $K_n = \omega(1)$ and $p_n p_{s,q} = p_{e,q} \leq \frac{2\ln n}{n}$, the proof of (\ref{prfhpnkn})
becomes evident by
\begin{align}
& \textrm{L.H.S. of (\ref{prfhpnkn})}
  \nonumber  \\
&  ~ \leq  \binom{m}{2} p_n \cdot \mathbb{P}[\mathcal {T}_m^{*} \in
\mathbb{T}_m \setminus \mathbb{T}_m^{(0)}] + \frac{q}{K_n} \cdot
p_n \cdot \mathbb{P}[\mathcal {T}_m^{*} \in \mathbb{T}_m^{(0)}] \nonumber  \\
& ~ \leq m^2 /2   \cdot  p_n \cdot  m^2 p_{s,q} / 2 +  \frac{q}{K_n} \nonumber  \\
& ~ \leq  m^4 n^{-1}\ln n / 2 + o(1) \nonumber  \\
& ~\to 0,\textrm{ as }n \to \infty. \nonumber
\end{align}

\subsection{Establishing (\ref{prob_MmMm_sim})}

We have
\begin{align}
&\mathbb{P} \big[ \big( \mathcal {M}_m   =  \mathcal{M}_m^{(0)}
\big) \mathlarger{\cap} \big( \mathcal {T}_m
  \in \mathbb{T}_m^{(0)} \big) \big]
 \nonumber  \\
& = \sum_{\mathcal {T}_m^{*} \in \mathbb{T}_m^{(0)}}  \Big\{
\mathbb{P} \big[ \mathcal {T}_m   =  \mathcal {T}_m^{*} \big]
\mathbb{P} \big[ \big( \mathcal {M}_m   =  \mathcal{M}_m^{(0)} \big)
\boldsymbol{\mid} \big( \mathcal {T}_m   =   \mathcal {T}_m^{*}
\big)\big] \Big\},\nonumber
\end{align}
where $\mathbb{P} \big[ \big( \mathcal {M}_m = \mathcal{M}_m^{(0)}
 \big) \boldsymbol{\mid} \big( \mathcal {T}_m
= \mathcal {T}_m^{*} \big)\big]$ as given by (\ref{pMmexpr}) equals%
%
%
\begin{align}
 &f\big(n - m , \mathcal{M}_m^{(0)}\big)
\mathbb{P} [w \in M_{0^m}
 \boldsymbol{\mid} 
\mathcal {T}_m  = \mathcal {T}_m^{*} ]^{n-m-hm}\nonumber  \\
& ~~\times \prod_{i=1}^{m}\big\{ \mathbb{P}[w \in
M_{0^{i-1}, 1, 0^{m-i}} \boldsymbol{\mid}
\mathcal {T}_m = \mathcal {T}_m^{*} ]^{h} \big\},\label{eqn_probMm}
\end{align}
with $f\big(n-m , \mathcal{M}_m^{(0)}\big)$ computed in
(\ref{eqn_f00}). For $\mathcal {T}_m^{*} \in \mathbb{T}_m^{(0)}$,
from $|S_{ij}^{*}| < q$ and (\ref{eq_evalprob_2_qcmp}) in Lemma
\ref{lem_evalprob_qcmp}, we derive
\begin{align}
&\mathbb{P}\big[w \in M_{0^{i-1}, 1, 0^{m-i}} \boldsymbol{\mid}
\mathcal {T}_m = \mathcal {T}_m^{*} \big] \nonumber  \\ & \quad \geq
p_{e, q} \bigg[ 1 - (q+2)!m{(p_{e, q})}^{\frac{1}{q}} - \frac{qp_n
}{K_n} \bigg] \label{eqn_wM0} .
\end{align}
Substituting (\ref{eqn_f00}) (\ref{eqn_wM0}) above and
(\ref{eq_evalprob_3_qcmp}) in Lemma \ref{lem_evalprob_qcmp} into
(\ref{eqn_probMm}), and using $p_{e,q} = o(1)$ and $K_n =
\omega(1)$, we conclude that
\begin{align}
&\mathbb{P} \big[ \big( \mathcal {M}_m   =  \mathcal{M}_m^{(0)}
\big) \mathlarger{\cap} \big( \mathcal {T}_m
  \in \mathbb{T}_m^{(0)} \big) \big]
 \nonumber  \\
& \quad \geq  \mathbb{P}[\mathcal {T}_m
  \in \mathbb{T}_m^{(0)}] \cdot  (h!)^{-m}n^{hm} \cdot [1-o(1)]
  \nonumber  \\
 & \quad \quad \times  (1 - m p_{e,q})^{n-m-hm}  {p_{e,q}}^{hm}   \nonumber  \\
 & \quad \quad \times \bigg[ 1 - (q+2)!m{(p_{e,q})}^{\frac{1}{q}} - \frac{qp_n
}{K_n} \bigg]^{hm}
 \nonumber  \\
 &  \quad \sim (h!)^{-m} (n p_{e,q})^{hm} e^{-m n p_{e,q}}. \nonumber
\end{align}

\section{Experimental Results} \label{sec:expe}

To confirm the theoretical results, we
now provide experiments in the non-asymptotic regime;
i.e., when parameter values are set according to real-world
sensor network scenarios. As we will see, the
experimental observations are in agreement with
 our theoretical findings.



 \begin{figure}[!t]
  \centering
 \includegraphics[height=0.236\textwidth]{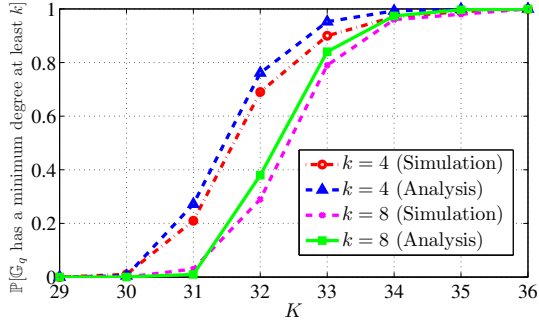}
\vspace{-10pt}\caption{A plot of the probability that graph
$\mathbb{G}_q(n,K,P,p)$ has a minimum node degree at least $k$ as a
function of $K$ for
         $k=4$ and $k=8$
         with $n=2,000$, $q=2$, $P=10,000$, and $p=0.8$.
         }
\label{f5}
\end{figure}

  \begin{figure}[!t]
  \centering
 \includegraphics[width=0.49\textwidth]{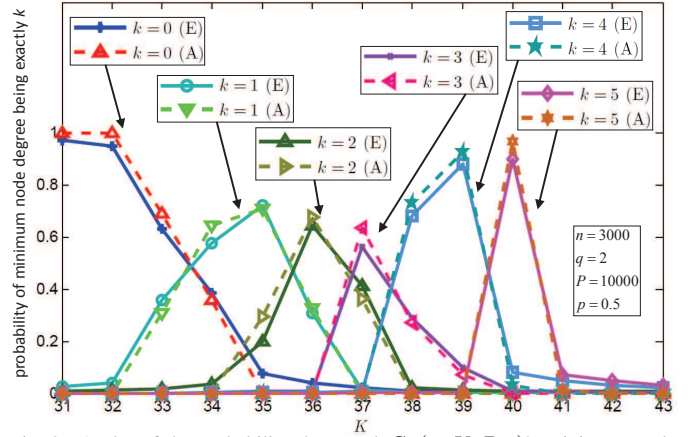} \vspace{-24pt}
\caption{A plot of the probability that graph
$\mathbb{G}_q(n,K,P,p)$'s minimum node degree equals $k$ exactly as a
function of $K$ for
         $k=0,1,2,3,4,5$
         with $n=3000$, $q = 2$, $P=10000$, and $p=0.5$. \vspace{-5pt}
         }
\label{fig5}
\end{figure}


\begin{figure}[!t]
  \centering
 \includegraphics[height=0.352\textwidth]{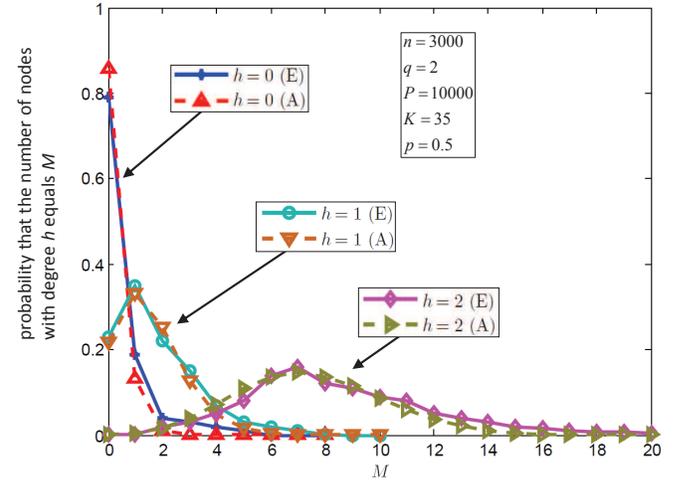} \vspace{-10pt}
\caption{A plot of the probability
 distribution for the number of nodes with
degree $h$ for $h=0,1,2 $ in graph $\mathbb{G}_q(n,K,P,p)$ with
$n=3000$, $q = 2$, $P=10000$, $K = 35 $ and $p = 0.5 $. \vspace{-5pt}
  } \label{fig2}
\end{figure}


In Figure \ref{f5}, we depict the probability that graph
$\mathbb{G}_q(n,K,P,p)$ has a minimum node degree at least $k$ from
both the simulation and the analysis, for $k = 4,8$ and
$K$ varying from 29 to 36 (we set $n=2,000$, $q = 2$ and $P=10,000$ and $p=0.8$).
On the one hand, for the experimental curves in all figures, we generate $2,000$ independent
samples of $\mathbb{G}_q(n,K,P,p)$ given a parameter set and record the
 count (out of a possible $2,000$) that the minimum degree
of graph $\mathbb{G}_q(n,K,P,p)$ is no less than $k$. Then the empirical probabilities are
obtained by dividing the
 counts by $2,000$. On the other hand, we approximate the analytical curves of Figure \ref{f5} by the asymptotic
results as explained below. First, we
compute the corresponding probability of $p_{e,q}$ in
$\mathbb{G}_q(n,K,P,p)$ through $p_{e,q} = p \cdot \sum_{u=q}^{K}
\big[{\binom{K}{u}\binom{P-K}{K-u}}\big/{\binom{P}{K}}\big] $ given
(\ref{psq2cijFC3})  and $P > 2 K$. Then
 we determine $\alpha$ by (\ref{peq1sbsc}) (we write $\alpha_n$ as
$\alpha $ here as $n$ is fixed); i.e., $p_{e, q}  = \frac{\ln  n + {(k-1)} \ln \ln n + {\alpha}}{n}$. Then given Remark \ref{thm:exact_qcomposite-rem} after Theorem \ref{thm:exact_qcomposite}, we
plot the analytical curves by considering that the minimum degree of
$\mathbb{G}_q(n,K,P,p)$ is at least $k $ with probability $e^{-
\frac{e^{-\alpha}}{(k-1)!}}$. The observation that the simulation and the analytical curves in Figure \ref{f5} are close is in accordance with
Theorem \ref{thm:exact_qcomposite}.

 In Figures \ref{fig5} and \ref{fig2}, the curves with legends labelled ``(E)'' are \emph{experimental} curves produced from experiments, while the curves with legends labelled ``(A)'' are \emph{analytical} curves generated from theoretical analysis.
In Figure \ref{fig5}, we depict the probability that graph
$\mathbb{G}_q(n,K,P,p)$'s minimum node degree equals $k$ exactly as a
function of $K$ for
         $k=0,1,2,3,4,5$. We set $n=3000$, $q = 2$, and $P=10000$, and $p=0.5$. For the experimental curves, we
generate $2000$ independent samples of graph
$\mathbb{G}_q(n,K,P,p)$ and record the count that the minimum degree
of graph $\mathbb{G}_q(n,K,P,p)$ is exactly $k$; and the
empirical probability of $\mathbb{G}_q(n,K,P,p)$ having a minimum
degree of $k$ is derived by averaging over the $2000$
experiments. The analytical curves are produced as follows. First, we
compute the corresponding probability of $p_{e,q}$ in
$\mathbb{G}_q(n,K,P,p)$ through the aforementioned expression $p_{e,q} = p \cdot \big\{1- \sum_{u=0}^{q-1}
\big[{\binom{K}{u}\binom{P-K}{K-u}}\big/{\binom{P}{K}}\big]\big\} $. Then we select $\ell^*$ such that $\big|p_{e, q} - \frac{\ln  n + (\ell-1) \ln \ln n }{n}\big|$ is minimized for integer $\ell$ (i.e., $\ell^*=\argmin_{\textrm{integer }\ell}\big|p_{e, q} - \frac{\ln  n + (\ell-1) \ln \ln n }{n}\big|$) and further define $\gamma^*$ such that $p_{e, q} = \frac{\ln  n + (\ell-1) \ln \ln n +\gamma^*}{n}$. Given Remark \ref{thm:exact_qcomposite-more-fine-grained-rem} after Theorem \ref{thm:exact_qcomposite-more-fine-grained}, we plot the analytical curves by considering that i) if $\ell^* > 0$, then $\bP{\delta =  k}$ equals $e^{- \frac{e^{-\gamma ^*}}{(\ell^*-1)!}}$ for $k=\ell^*$,  equals $1-e^{- \frac{e^{-\gamma ^*}}{(\ell^*-1)!}}$ for $k=\ell^*-1$, and equals $0$ for $k\neq\ell^*$ and $k\neq\ell^*-1$, and ii) if $\ell^* \leq 0$, then $\bP{\delta =  k}$ equals $1$ for $k=0$, and equals $0$ for $k\neq0$.
  The observation that the curves
generated from the experimental and the analytical curves are close to
each other confirms the result on the distribution of the minimum   degree
in Theorem \ref{thm:exact_qcomposite-more-fine-grained}.


 In Figure \ref{fig2}, we plot the
probability
 distribution for the number of nodes with degree $h$
  in graph $\mathbb{G}_q(n,K,P,p)$ for $h=0,1,2$ from both the
experiments and the analysis. We set $n=3000$, $q = 2$, $K = 35 $, $P=10000$, and $p=0.5$. On the one hand, for the experiments, we generate $2000$ independent
samples of $\mathbb{G}_q(n,K,P,p)$ and record the
 count (out of a possible $2000$) that the number
of nodes with degree $h$ for each $h$ equals a particular
non-negative number $M$. Then the empirical probabilities are
obtained by dividing the
 counts by $2000$. On the other hand, we approximate the analytical curves by the asymptotic
results as explained below. In Theorem \ref{thm:exact_qcomposite2}, we establish that the number of nodes
in $\mathbb{G}_q(n,K_n,P_n,p_n)$ with degree $h$
approaches to a Poisson distribution with mean $\lambda_{n,h}=n
(h!)^{-1}(n p_{e,q})^h e^{-n p_{e,q}}$ as $n \to \infty$. We derive
$\lambda_{n,h}$ by computing the corresponding probability of $p_{e,q}$
in $\mathbb{G}_q(n,K,P,p)$ through $p_{e,q} = p \cdot \big\{1- \sum_{u=0}^{q-1}
\big[{\binom{K}{u}\binom{P-K}{K-u}}\big/{\binom{P}{K}}\big]\big\} $ as explained above. Then for each $h$, we plot
a Poisson distribution with mean $\lambda_{n,h}$ as the curve
corresponding to the analysis. In Figure \ref{fig2}, we observe that the curves generated
from the experiments and those obtained by the analysis are close to
each other, confirming the result on asymptotic Poisson distribution
in Theorem \ref{thm:exact_qcomposite2}.

\section{Related Work} \label{related}

Erd\H{o}s and R\'{e}nyi \cite{citeulike:4012374} propose the random graph model $G(n,p_n)$ defined on
a node set with size $n$ such that an edge between any two nodes
exists with probability $p_n$ \emph{independently} of all other
edges. For graph $G(n,p_n)$, Erd\H{o}s and R\'{e}nyi
\cite{citeulike:4012374} derive the asymptotically exact
probabilities for connectivity and the property that the minimum degree
is at least $1$, by proving first that the number of isolated nodes
converges to a Poisson distribution as $n \to \infty$. Later, they
extend the results to general $k$ in \cite{erdos61conn}, obtaining
the asymptotic Poisson distribution for the number of nodes with any
degree and the asymptotically exact probabilities for
$k$-connectivity and the event that the minimum degree is at least
$k$, where $k$-connectivity is defined as the property that the
network remains connected in spite of the removal of any $(k-1)$
nodes.

Recall that graph $\mathbb{G}_q(n, K_n, P_n)$ models the topology of the $q$-composite key predistribution scheme \cite{ANALCO,farrell2015hyperbolicity,6875009}.
For graph $\mathbb{G}_q(n, K_n, P_n)$, Bloznelis \emph{et al.}
\cite{Rybarczyk} demonstrate that a connected component with at at
least a constant fraction of $n$ emerges asymptotically when
probability $p_{e,q}$ exceeds $1/n$. Recently, still for $G_q(n,
K_n, P_n)$, Bloznelis \cite{bloznelis2013} establishes the
asymptotic Poisson distribution for the number of nodes with any
degree. Our results in Theorem \ref{thm:exact_qcomposite2} by
setting $p_n$ as $1$ imply his result; in particular, the result
that he obtains is a special case of property (a) in our Theorem
\ref{thm:exact_qcomposite2}.

Ya\u{g}an \cite{yagan_onoff} presents
zero-one laws in graph $\mathbb{G}_1$ (our graph $\mathbb{G}_q$ in
the case of $q=1$) for connectivity and for the property that the
minimum degree is at least $1$. Zhao \emph{et al.} extend Ya\u{g}an's results to
general $k$ for $\mathbb{G}_1$ in \cite{ZhaoYaganGligor,ISIT}.
Our results in this paper apply to general $q$, yet the corresponding results for
$q=1$ are already stronger than those in
\cite{yagan_onoff,ISIT,ZhaoYaganGligor}.


Krishnan \emph{et al.} \cite{ISIT_RKGRGG} and Krzywdzi\'{n}ski and
Rybarczyk \cite{Krzywdzi} describe results for the probability of
connectivity asymptotically converging to 1 in WSNs employing the
$q$-composite key predistribution scheme with $q=1$ (i.e., the
Eschenauer-Gligor key predistribution scheme), not under the on/off
channel model but under the well-known disk model
\cite{ISIT_RKGRGG,Krzywdzi,ZhaoAllerton,6909183}, where nodes are distributed
over a bounded region of a Euclidean plane, and two nodes have to be
within a certain distance for communication. Simulation results in
our work \cite{ZhaoYaganGligor} indicate that for WSNs under the key
predistribution scheme with $q=1$, when the on-off channel model is
replaced by the disk model, the performances for $k$-connectivity
and for the property that the minimum degree is at least $k$ do not
change significantly.

\section{Conclusion and Future Work}
\label{sec:Conclusion}

In this paper, we analyze  topological properties  in WSNs operating under the $q$-composite key
predistribution scheme with on/off channels. Experiments are
shown to be in agreement with our theoretical findings. A future research direction  is to consider communication models different from the on/off channel model.

\appendix

\subsection{Additional Lemmas}


\begin{lem} \label{lem_eval_psq}
The following two properties hold, where $p_{s,q} $ denotes the probability that two  nodes in graph $\mathbb{G}_q$
share at least $q$ keys:
\begin{itemize}
\item[(i)] If $K_n = \omega(1)$ and $\frac{{K_n}^2}{P_n} = o(1)$, then\\$p_{s,q}
= \frac{1}{q!} \big( \frac{{K_n}^2}{P_n} \big)^{q} \times [1\pm o(1)]$; i.e., $p_{s,q}
\sim \frac{1}{q!} \big( \frac{{K_n}^2}{P_n} \big)^{q}$.\vspace{3pt}
\item[(ii)] If $K_n = \omega(\ln n)$ and $\frac{{K_n}^2}{P_n} = o\big(\frac{1}{\ln n}\big)$, then\\$p_{s,q}
= \frac{1}{q!} \big( \frac{{K_n}^2}{P_n} \big)^{q} \times [1\pm o\big(\frac{1}{\ln n}\big)]$.
\end{itemize}
\end{lem}

\begin{lem} \label{lem_evalprob_qcmp}
In graph $\mathbb{G}_q$, with $p_{e,q} $ denoting the probability that two
distinct nodes have a secure link in between, for any $\mathcal
{T}_m^{*} = (S_1^{*} ,  S_2^{*}  , \ldots, S_m^{*} ) \in
\mathbb{T}_m$ and any node $w \in \overline{\mathcal {V}_m} $, we
obtain
\begin{align}
 &  \mathbb{P} [w \in M_{0^m} \boldsymbol{\mid} \mathcal {T}_m =
\mathcal {T}_m^{*} ] \geq 1 - m p_{e, q} ,
\label{eq_evalprob_3_qcmp}
\end{align}
and for any $i = 1,2,\ldots,m $,
\begin{align}
 & \mathbb{P}\big[w \in M_{0^{i-1}, 1, 0^{m-i}} \boldsymbol{\mid}
\mathcal {T}_m = \mathcal {T}_m^{*} \big] \leq p_{e, q};
\label{eq_evalprob_1_qcmp}
\end{align}
and if $\frac{{K_n}^2}{P_n} = o(1)$, the following
(\ref{eq_evalprob_4_qcmp}) and (\ref{eq_evalprob_2_qcmp}) hold:
\begin{align}
& \mathbb{P} [w \in M_{0^m} \boldsymbol{\mid} \mathcal
{T}_m = \mathcal {T}_m^{*} ]  \nonumber \\
& \quad \leq e^{- m p_{e, q} +  (q+2)! \binom{m}{2}{(p_{e,
q})}^{\frac{q+1}{q}} +
  \frac{p_{e, q} p_n}{K_n}\sum_{1\leq i <j \leq m}|S_{i j}^{*}|},
   \label{eq_evalprob_4_qcmp}
\end{align}
and for any $i = 1,2,\ldots,m $,
\begin{align}
& \mathbb{P}\big[w \in M_{0^{i-1}, 1, 0^{m-i}} \boldsymbol{\mid}
\mathcal {T}_m = \mathcal {T}_m^{*} \big]  \nonumber   \\
& \quad \geq p_{e, q} \bigg[ 1 \hspace{-.5pt}-\hspace{-.5pt}
(q\hspace{-.5pt}+\hspace{-.5pt}2)!m{(p_{e,q})}^{\frac{1}{q}}
\hspace{-.5pt}-\hspace{-.5pt} \frac{p_n}{K_n}
\hspace{-.5pt}\sum_{j\in\{1,2,\ldots,m\}
\setminus\{i\}}\hspace{-.5pt} |S_{i j}^{*}| \bigg],
 \label{eq_evalprob_2_qcmp}
\end{align}
where $S_{ij}^{*} = S_{i}^{*} \cap S_{j}^{*}$.

\end{lem}

\begin{lem} \label{lem_psukn_qcmp}
In graph $\mathbb{G}_q$, if
$\frac{{K_n}^2}{P_n} = o(1)$, then for any three distinct nodes
$v_i, v_j$ and $v_t$ and for
any $u = 0, 1, \ldots, K_n$, we obtain that with sufficiently large
$n$,
\begin{align}
  \mathbb{P}[({\Gamma}_{i t} \cap {\Gamma}_{j t} \boldsymbol{\mid}
(|S_{ij}| = u)]  \leq \frac{ p_{s,q} u}{K_n}+
(q+2)! \cdot  ({p_{s,q})}^{\frac{q+1}{q}} . \nonumber
\end{align}
\end{lem}


Due to space limitation, we provide the proofs of Lemmas \ref{lem_eval_psq}, \ref{lem_evalprob_qcmp}, \ref{lem_psukn_qcmp} in Appendices B, C, D of the full version \cite{full}.

\normalsize


\end{document}